\documentclass[nofootinbib,showpacs, aps,prb,twocolumn,preprintnumbers ,amsmath, amssymb, superscriptaddress, aps]{revtex4-2}
\usepackage{graphicx}
\usepackage{dcolumn}
\usepackage{bm}
\usepackage{hyperref}
\usepackage{xcolor}
\usepackage{caption}
\usepackage{subcaption}
\usepackage[linesnumbered,ruled]{algorithm2e}
\usepackage{booktabs}
\usepackage{multirow}
\usepackage{rotating}
\usepackage{makecell}
\usepackage{enumitem}

\SetAlFnt{\small}
\SetAlCapFnt{\small}
\SetAlCapNameFnt{\small}

\begin{document}

\title{Fourier Neural Operators for Composition-Driven Crystal Structure Discovery}

\author{Zhijie Yu}
\affiliation{School of Mathematical Sciences, University of Science and Technology of China, Hefei 230026, China}
\affiliation{Suzhou Institute for Advanced Research, University of Science and Technology of China, Suzhou 215213, China}

\author{Jingyu Li}
\affiliation{School of Artificial Intelligence and Data Science, University of Science and Technology of China, Hefei 230026, China}
\affiliation{Suzhou Institute for Advanced Research, University of Science and Technology of China, Suzhou 215213, China}

\author{Yang Huang}
\email{hyhy123@ustc.edu.cn}
\affiliation{University of Science and Technology of China, Hefei 230026, China}
\affiliation{Suzhou Institute for Advanced Research, University of Science and Technology of China, Suzhou 215213, China}

\author{Jingrun Chen}
\email{jingrunchen@ustc.edu.cn}
\affiliation{School of Mathematical Sciences, University of Science and Technology of China, Hefei 230026, China}
\affiliation{Suzhou Institute for Advanced Research, University of Science and Technology of China, Suzhou 215213, China}

\date{\today}

\begin{abstract}
  Crystalline materials discovery is essential for energy, electronics, and catalysis, but the vast chemical and structural space makes exhaustive screening infeasible. Existing voxel-based methods are limited by the local receptive fields of three-dimensional convolutional neural networks and the posterior collapse of high-dimensional variational autoencoders. Here, we develop a Fourier Neural Operator (FNO)-based crystal-field solver that maps a prescribed chemical formula and lattice parameters to periodic number-density and electron-density fields. By operating on global Fourier modes, the solver captures long-range correlations in periodic crystal fields beyond conventional local convolutions. Building on this solver, we construct a coupled generation-solving framework in which a conditional variational autoencoder generates diverse candidate lattice parameters in a low-dimensional basis-coefficient space, followed by density-field prediction and atomic reconstruction through peak detection, position optimization, and weight optimization. The reconstructed structures are further screened using voxel-level filtering and machine-learning interatomic-potential relaxation. The framework generates structures not present in the Materials Project across 104 chemical formulas from the test set. Beyond the test set, the framework was applied to 3,000 chemical formulas absent from the training, validation, and test splits and from three reference databases; it produced 29,910 valid structures, of which 24,582 converged under DPA3 relaxation and 18,363 exhibit formation energies below $-0.2$~eV atom$^{-1}$. By extending Fourier neural operators to periodic crystal fields and coupling them with composition-conditioned lattice generation, our approach provides a scalable route to crystal structure discovery from prescribed chemical compositions.
\end{abstract}
\maketitle

\section{Introduction}

The discovery of novel crystalline materials is fundamental to advancing technologies in energy storage, electronics, catalysis, and quantum information \cite{Lewis2007, Magee2012, Snyder2010}. However, the combinatorial space of chemical elements, stoichiometries, and crystal structures is astronomically large, rendering exhaustive experimental or computational screening infeasible \cite{Oganov2019}. Traditional computational approaches to crystal structure prediction (CSP)---such as genetic algorithms \cite{Oganov2006, Falls2021}, particle swarm optimization \cite{Wang2010}, random structure searching \cite{Pickard2011}, and minima hopping \cite{Goedecker2004}---explore the potential energy surface by iteratively evaluating candidate structures with density functional theory (DFT) calculations \cite{Kresse1996}. While these methods have enabled the discovery of numerous novel materials \cite{Yamashita2021}, they remain computationally expensive and scale poorly with system size, as each candidate requires explicit energy evaluation.

In recent years, deep generative models have emerged as a powerful alternative that learns the underlying distribution of crystal structures from large databases, bypassing the computationally intensive iterative search inherent to traditional CSP methods \cite{DeBreuck2025Review, Li2024Review}. By capturing the complex, multimodal distribution $p(\mathcal{M})$ over material structures, these models can directly propose novel, plausible crystal structures without \textit{a priori} constraints on chemistry or stoichiometry. Moreover, many generative frameworks support conditional generation $p(\mathcal{M} \mid c)$, enabling targeted discovery of materials with desired chemical compositions or functional properties \cite{Xie2022CDVAE, Zeni2025MatterGen}.

A wide range of generative architectures have been adapted for crystal generation, including variational autoencoders (VAEs) \cite{Kingma2014VAE}, generative adversarial networks (GANs) \cite{Goodfellow2014GAN}, transformers \cite{Vaswani2017Attention}, normalizing flows \cite{Rezende2015Flows, Dinh2017RealNVP}, diffusion models \cite{Ho2020DDPM, Song2021SDE, Jiao2024DiffCSP}, and fine-tuned large language models \cite{Touvron2023Llama2, Gruver2024CrystalTextLLM}. These models operate on various invertible representations of crystal structures, such as point clouds with lattice matrices $(\mathbf{A}, \mathbf{X}, \mathbf{L})$ \cite{Jiao2024DiffCSP, Zeni2025MatterGen}, voxel grids \cite{Noh2019iMatGen, Hoffmann2019}, graphs \cite{Xie2022CDVAE}, reciprocal-space descriptors \cite{Ren2022FTCP}, and Wyckoff-position-based representations \cite{Cao2024CrystalFormer, DeBreuck2025MatraGenoa, Jiao2024DiffCSPPlus}. 
The choice of representation fundamentally affects how symmetry, periodicity, and atomic details are captured, and many state-of-the-art models employ multiple representations simultaneously \cite{DeBreuck2025Review}. Recent reviews provide systematic comparisons of these approaches and highlight diffusion-based methods such as MatterGen as particularly promising for conditional crystal generation \cite{Li2024Review, DeBreuck2025Review}.

Voxel-based methods are attractive for crystal generation because they represent structures as continuous density fields on regular three-dimensional grids, preserving geometric information while avoiding explicit atomic connectivity and atom-ordering ambiguities. Most existing approaches use 3D convolutional neural network (CNN)-based variational autoencoders (VAEs), with iMatGen \cite{Noh2019iMatGen} providing an early demonstration for inorganic materials, followed by U-Net/VAE \cite{Hoffmann2019} and conditional-VAE approaches \cite{Court2020}. Recent work has extended this paradigm to diffusion models operating on 3D voxelized representations, including CrysTens-3D, which employs a CNN-based U-Net architecture with atom feature encoding for crystal structure generation \cite{Ramachandran2025CrysTens}. These methods encode density fields into low-dimensional latent spaces from which new structures can be generated.

However, this paradigm faces two fundamental limitations. First, the local nature of 3D convolutions limits their ability to capture the long-range periodic correlations characteristic of crystal density fields. While Fourier representations have been explored for crystalline materials \cite{Ren2022FTCP}, their integration with voxel-based generative models remains unexplored. The Fourier Neural Operator (FNO) \cite{Li2021FNO} offers a principled framework for learning mappings between function spaces through spectral convolutions in Fourier space, naturally capturing global correlations while remaining discretization-invariant \cite{Duruisseaux2025FNO}. Second, the high dimensionality of voxel representations can cause the reconstruction loss to dominate VAE training, weakening KL regularization and promoting posterior collapse \cite{Bowman2016Posterior,Zhao2017Posterior,Higgins2017BetaVAE}. Together, these limitations restrict both the representation of global crystal structure and the diversity of generated candidates.

To address these limitations, we develop a coupled generation-solving workflow that separates lattice generation from crystal density-field prediction. A conditional variational autoencoder (CVAE) \cite{Kingma2014VAE, Sohn2015CVAE} generates lattice parameters conditioned on the chemical formula, while the proposed CrystalFNO Solver, based on the Fourier Neural Operator (FNO) \cite{Li2021FNO}, predicts the corresponding number-density and electron-density fields. This separation avoids direct generative modeling in the high-dimensional voxel space while exploiting global spectral convolutions to capture long-range correlations in periodic crystal fields. The predicted fields are subsequently converted into atomic structures and screened through voxel filtering, Deep Potential Atomistic (DPA) relaxation \cite{Zhang2018DP, Wang2018DP}. Across 104 chemical formulas, the workflow generates diverse and structurally valid crystal candidates, demonstrating an efficient strategy for composition-driven crystal discovery. To our knowledge, CrystalFNO represents the first application of Fourier Neural Operators as crystal-field solvers within a voxel-based crystal structure generation framework.

\section{Method}
\begin{figure*}[thpb]
  \centering
  \includegraphics[width=0.95\textwidth]{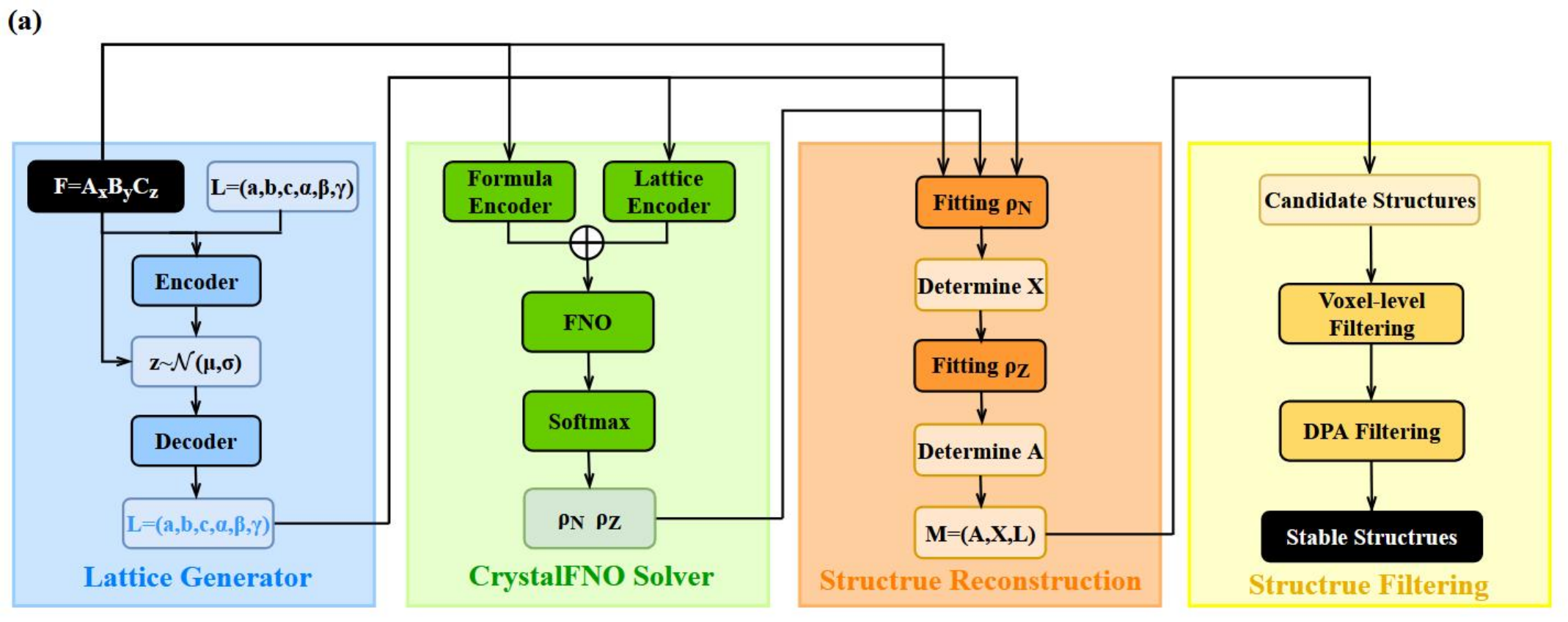}\\
  \includegraphics[width=0.30\textwidth]{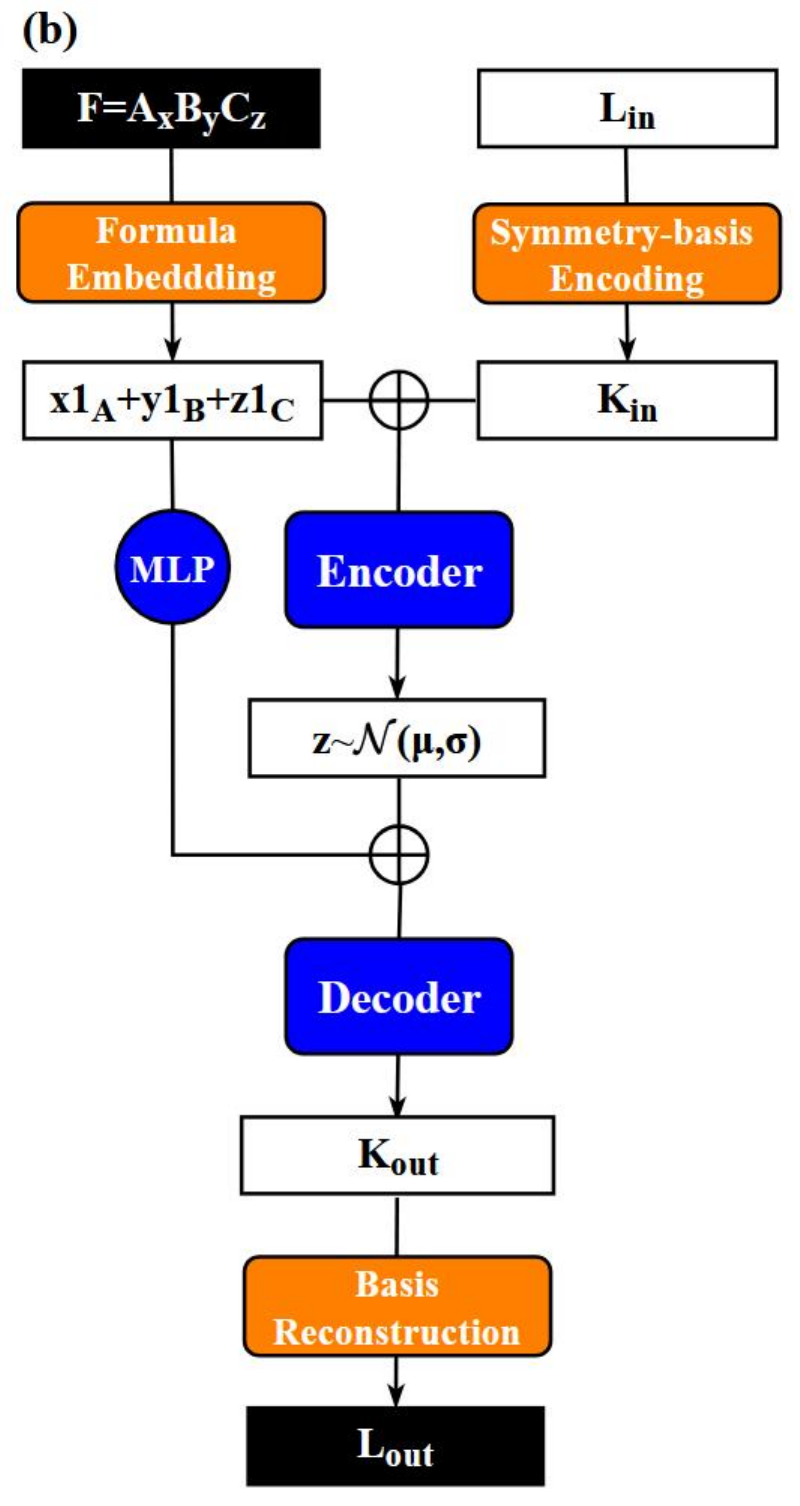}
  \includegraphics[width=0.67\textwidth]{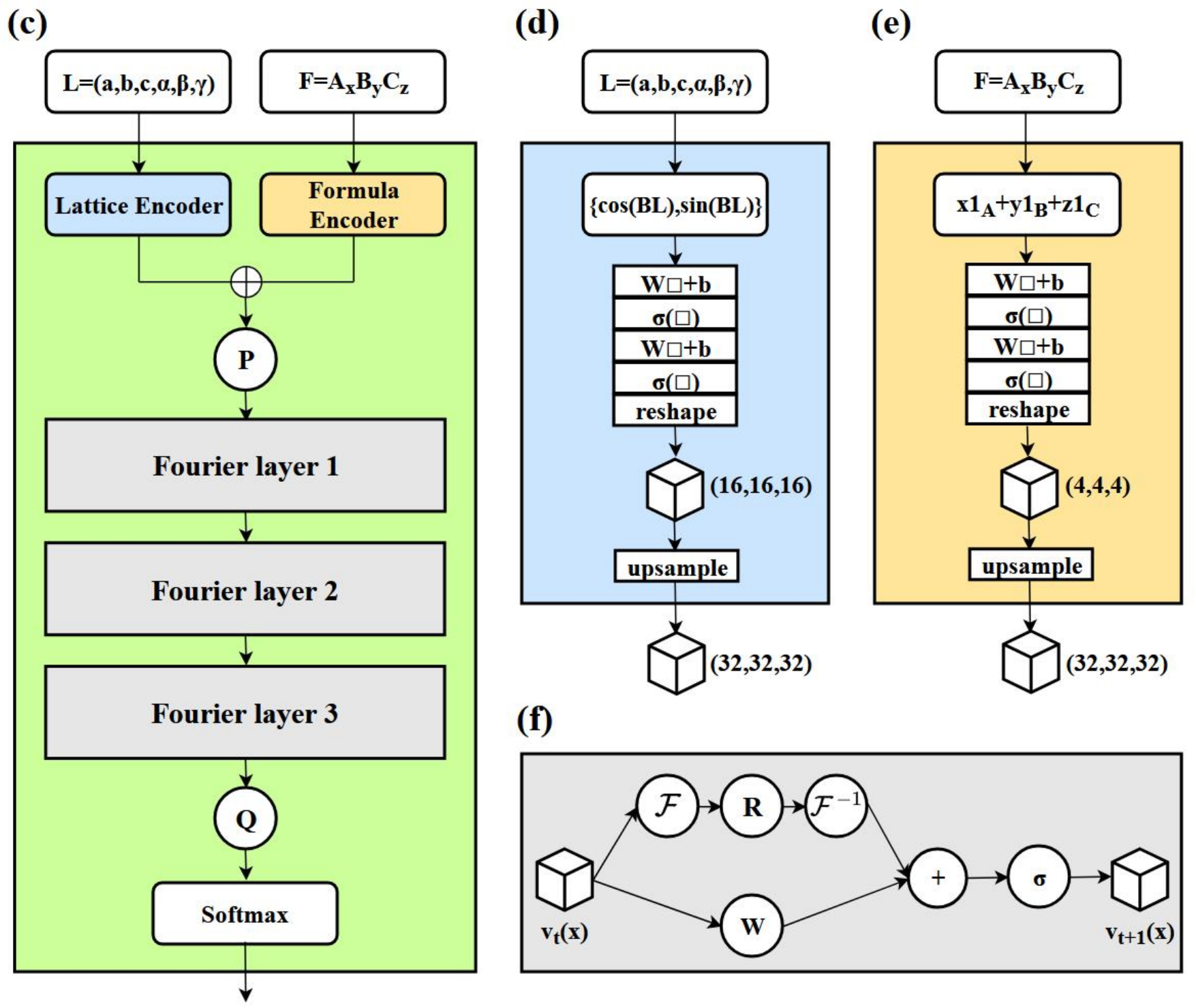}
  \caption{
    Overview of the proposed coupled generation-solving workflow for composition-driven crystal structure discovery.
    (a) Overall workflow, from lattice generation and density-field prediction to structure reconstruction and sequential screening.
    (b) Conditional variational autoencoder for composition-conditioned lattice generation.
    (c) CrystalFNO Solver for predicting the number-density field $\rho_{N}$ and electron-density field $\rho_Z$. Panel (c1) shows the overall solver; panels (d)--(f) detail the formula encoder, the lattice encoder, and the Fourier layer, respectively.
    (d) Lattice-parameter encoder that maps the lattice parameters $(a,b,c,\alpha,\beta,\gamma)$ to Fourier-encoded feature volumes.
    (e) Chemical-formula encoder that encodes the chemical formula $F=A_xB_yC_z\cdots$ into composition-weighted elemental feature volumes.
    (f) Fourier layer with parallel frequency-domain and spatial branches for capturing long-range periodic correlations and local atomic-scale features.
}
\label{fig:overall_pipeline}
\end{figure*}

The overall workflow is illustrated in Fig.~\ref{fig:overall_pipeline}~(a). Given a chemical formula $F=A_xB_yC_z\cdots$, the framework first generates candidate lattice parameters $L=(a,b,c,\alpha,\beta,\gamma)$, followed by prediction of the corresponding number-density field $\rho_N$ and electron-density field $\rho_Z$. These density fields are then used to reconstruct the crystal structure $M=(A,X,L)$, where $A$ denotes the elemental species, $X$ the fractional atomic coordinates, and $L$ the lattice parameters. The reconstructed structures are subsequently screened through voxel-level filtering and structural relaxation.

\subsection{Lattice Generator}
We use a conditional variational autoencoder (CVAE) \cite{Kingma2014VAE} to generate lattice parameters from the chemical formula, as illustrated in Fig.~\ref{fig:overall_pipeline}(b). Rather than generating the conventional lattice parameters $(a,b,c,\alpha,\beta,\gamma)$ directly, the CVAE predicts a six-dimensional coefficient vector $\mathbf{k}=[k_1,k_2,\ldots,k_6]$ obtained through a symmetric basis decomposition. This parameterization ensures a positive-definite lattice metric and allows crystal-system-specific linear constraints to be imposed on $\mathbf{k}$. The basis decomposition follows the symmetry constraints defined in the International Tables for Crystallography \cite{InternationalTables} and the formulation adopted in DiffCSP++ \cite{Jiao2024DiffCSPPlus}. 
Further architectural details and training objectives are provided in Appendix~\ref{app:basis_comp} and \ref{app:cvae}.

\subsection{CrystalFNO Solver}
\paragraph{Voxel Representation} Under periodic boundary conditions, we define a Gaussian kernel function. Let $x \in [0,1)^3$ be the voxel center coordinates and $u_k \in [0,1)^3$ be the fractional coordinates of the $k$-th atom. The Cartesian displacement vector satisfying the minimum image convention is defined as
\begin{equation}
d(x,u_k) = \left( x - u_k - \text{round}(x - u_k) \right) \cdot \bm{L},
\end{equation}
where $\text{round}(\cdot)$ denotes component-wise rounding to the nearest integer, and $\bm{L} \in \mathbb{R}^{3 \times 3}$ is the lattice matrix. The corresponding periodic Gaussian kernel is defined as
\begin{equation}
G(x - u_k;\sigma) = \frac{1}{(2\pi\sigma^2)^{3/2}} \exp\left( -\frac{\| d(x,u_k)\|^2}{2\sigma^2} \right).
\end{equation}
In this work, we set the Gaussian width to $\sigma = 0.5\ \text{\AA}$ and discretize the unit cell into a $32 \times 32 \times 32$ grid mesh.

Let the total number of atoms be $N$, and the atomic number of the $k$-th atom be $Z_k$. We define the number-density field $\rho_{N}$ and the electron-density field $\rho_Z$ as follows
\begin{align}
\rho_{N}(x_i) &= \frac{|\det \bm{L}|}{N} \sum_{k=1}^{N} G(x_i - u_k;\sigma),\\
\rho_Z(x_i) &= |\det \bm{L}| \sum_{k=1}^{N} z_k G(x_i - u_k;\sigma),
\end{align}
where $z_k = Z_k / \sum_{k=1}^{N} Z_k$ is the normalized coefficient.

Both fields are normalized such that their integrals over the unit cell are unity. The number-density field contains only positional information independent of element types, making it suitable for atom localization; the electron-density field simultaneously encodes both positional and elemental information, supporting subsequent element type assignment.

\paragraph{FNO-Based Density Solver} The architecture of the CrystalFNO Solver is illustrated in Fig.~\ref{fig:overall_pipeline}(c). The periodic nature of the crystal domain naturally aligns with the spectral representation in Fourier space. The solver comprises a lattice encoder, a formula encoder, and three Fourier layers, which together map the chemical formula and lattice parameters to normalized number-density $\rho_N$ and electron-density fields $\rho_Z$ on the $32\times32\times32$ voxel grid.

The lattice parameters $(a,b,c,\alpha,\beta,\gamma)$ are encoded using Fourier features:
\begin{equation}
\gamma(p)=\bigoplus_{k=1}^{m}\left[
\cos\left(\frac{2\pi k p}{T}\right),
\sin\left(\frac{2\pi k p}{T}\right)
\right],
\end{equation}
where $\oplus$ denotes vector concatenation, $m=15$ is the number of Fourier modes, and $T$ is the normalization constant. The encoded features are processed by a two-layer MLP, reshaped into a $16\times16\times16$ feature volume, and upsampled to $32\times32\times32$ [Fig.~\ref{fig:overall_pipeline}~(d)].

The chemical formula $F=A_xB_yC_z\cdots$ is encoded as a composition-weighted sum of 118-dimensional elemental one-hot vectors,
\begin{equation}
E(F)=x\mathbf{1}_A+y\mathbf{1}_B+z\mathbf{1}_C+\cdots,
\end{equation}
where $x,y,z$ are the corresponding atomic counts. The resulting vector is processed by a two-layer multilayer perceptron (MLP), reshaped into a $4\times4\times4$ feature volume, and upsampled to $32\times32\times32$ [Fig.~\ref{fig:overall_pipeline}~(e)].

The lattice and formula feature volumes are concatenated along the channel dimension to provide conditional features for the subsequent FNO layers. The FNO processor comprises three sequential Fourier layers, each with parallel frequency-domain and spatial branches [Fig.~\ref{fig:overall_pipeline}(f)]. In the frequency-domain branch, the features are transformed using a 3D fast Fourier transform (FFT), processed by learnable spectral convolution kernels, and transformed back using the inverse FFT (IFFT) to capture long-range periodic correlations in the crystal density field \cite{Li2021FNO}. In parallel, the spatial branch applies a $3\times3\times3$ 3D convolution to extract local atomic-scale features. The outputs of the two branches are summed element-wise and passed through a nonlinear activation, thereby integrating global periodic correlations with local structural information.

After hierarchical feature extraction through the three stacked Fourier layers, the fused 3D features are mapped to a single-channel output volume via the decoder, followed by a Softmax operation to normalize it into a valid probability density distribution [Fig.~\ref{fig:overall_pipeline}~(c)].

In this work, we train two independent FNO models for the number-density field $\rho_{N}$ and the electron-density field $\rho_Z$, respectively, with no parameter sharing. Let $\widehat{\rho}(x_i)$ denote the predicted density field and $\rho(x_i)$ denote the ground-truth density field. We adopt the $L^1$ loss as the optimization objective:
\begin{equation}
\mathcal{L} = \frac{1}{M} \sum_{i=1}^{M} \left| \widehat{\rho}(\mathbf{x}_i) - \rho(\mathbf{x}_i) \right|,
\end{equation}
where $M = 32768$ corresponds to the total number of voxels in the $32 \times 32 \times 32$ grid.

\subsection{Structure Reconstruction}
Given the predicted number-density field $\widehat{\rho}_{N}$ and electron-density field $\widehat{\rho}_Z$ from the CrystalFNO Solver, we reconstruct atomic positions and elemental species by solving three cascaded optimization problems.

\paragraph{Position assignment} Atomic positions correspond to local maxima of the number-density field $\widehat{\rho}_{N}$. We adopt a $3 \times 3 \times 3$ non-maximum suppression strategy to identify candidate voxels, with an adaptive minimum distance constraint $d_{\min} = c\sigma$ (initialized at $c=2$, decreased stepwise to $c_{\min}=0.1$ if the candidate count is insufficient). The top $N$ candidates are selected as the initial atomic positions $\{\boldsymbol{\mu}_k^{(0)}\}$.

With atomic weights fixed at $w_k = 1$, we optimize atomic positions by minimizing the discrepancy between the reconstructed and predicted density fields:
\begin{equation}
\min_{\{\boldsymbol{\mu}_k\}} \quad \mathcal{L}_{\mathrm{pos}} 
= \frac{1}{M} \sum_{i=1}^{M} \left| \rho_{\text{N-recon}}(\mathbf{x}_i) - \widehat{\rho}_{N}(\mathbf{x}_i) \right|,
\end{equation}
where the reconstructed density field is defined as
\begin{equation}
\rho_{\text{N-recon}}(\mathbf{x}_i) = \frac{|\det \mathbf{L}|}{N} \sum_{k=1}^{N} G(\mathbf{d}_{ki};\sigma).
\end{equation}
The optimization is performed via gradient descent, with positions projected back to $[0,1)^3$ after each iteration. The refined positions are denoted as $\{\boldsymbol{\mu}_k^{\text{opt}}\}$.

\paragraph{Element assignment} With atomic positions fixed, we optimize the weights $\{w_k\}$ to fit the electron-density field $\widehat{\rho}_Z$:
\begin{equation}
\min_{\{w_k\}} \quad \mathcal{L}_{\mathrm{weight}} 
= \frac{1}{M} \sum_{i=1}^{M} \left| \rho_{Z\text{-recon}}(\mathbf{x}_i) - \widehat{\rho}_Z(\mathbf{x}_i) \right|,
\end{equation}
where
\begin{equation}
\rho_{Z\text{-recon}}(\mathbf{x}_i) = |\det \mathbf{L}| \sum_{k=1}^{N} w_k G(\mathbf{d}_{ki};\sigma),
\end{equation}
subject to the constraints
\begin{equation}
w_k \geq 0, \quad \sum_{k=1}^{N} w_k = 1.
\end{equation}
The weights are initialized as $w_k^{(0)} = v_k^{\text{new}} / \sum_{j} v_j^{\text{new}}$, where $v_k^{\text{new}} = \widehat{\rho}_Z(\text{voxel}(\boldsymbol{\mu}_k^{\text{opt}}))$ are voxel values queried from the predicted density field. After optimization, elements are assigned to atoms in descending order of their weights, yielding the final reconstructed structure.

\subsection{Structure Filtering}
The reconstructed structures are filtered through two sequential stages: voxel-level screening and DPA relaxation. Candidates are first ranked by their consistency with the CrystalFNO Solver-predicted density fields, and then candidates with the lowest voxel-level $L^1$ loss are retained for DPA relaxation.

\paragraph{Voxel-Level Screening} For each candidate, the voxel-level loss is calculated as
\begin{equation}
\mathcal{L}_{\mathrm{vox}}=\frac{1}{M}\sum_{i=1}^{M}\left|\widehat{\rho}(\mathbf{x}_i)-\rho_{\mathrm{rec}}(\mathbf{x}_i)\right|,
\end{equation}
where $\widehat{\rho}$ and $\rho_{\mathrm{rec}}$ denote the CrystalFNO Solver-predicted and reconstructed density fields, respectively, and $M$ is the number of voxels. Candidates are ranked by $\mathcal{L}_{\mathrm{vox}}$, with the 10 lowest-loss structures retained.

\paragraph{DPA Relaxation Screening} The retained structures are relaxed using the DPA3 model \cite{Zhang2018DP,Wang2018DP}, optimizing atomic positions and lattice parameters until the atomic forces are below $0.005$~eV/\AA. A relaxed structure is retained if all of the following criteria are satisfied: (i) each lattice parameter changes by no more than $5\%$; (ii) the atomic RMSD is no greater than $0.3$~\AA; and (iii) the element match rate is $100\%$.

\section{Results and Discussion}

\subsection{CrystalFNO Solver Density Field Prediction}
The CrystalFNO Solver is trained and evaluated on the MP20 dataset derived from the Materials Project database \cite{Jain2013MP}, which contains 44,372 crystal structures spanning diverse chemical compositions and crystal symmetries. The dataset is divided into training, validation, and test sets containing 35,491, 4,439, and 4,442 structures, respectively. All structures are Niggli reduced \cite{Niggli1928} to provide a standardized unit-cell representation for lattice encoding and density-field construction.

The test set contains 4,442 structures comprising 85 elements and 142 space groups across all seven crystal systems, providing a diverse basis for evaluating the generalization of the CrystalFNO Solver. We first assess the accuracy of the predicted number-density and electron-density fields before evaluating structure reconstruction and subsequent filtering.

\begin{figure}[htbp]
\centering
\includegraphics[width=0.5\textwidth]{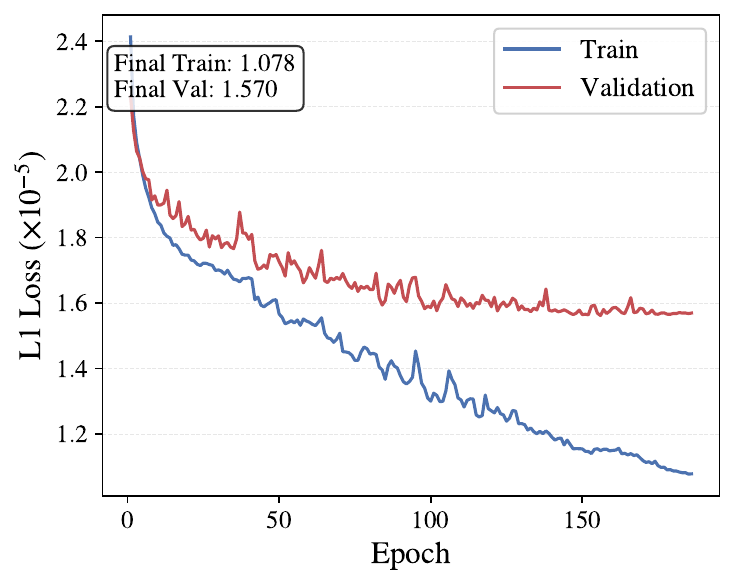}\\[0.5em]
\includegraphics[width=0.5\textwidth]{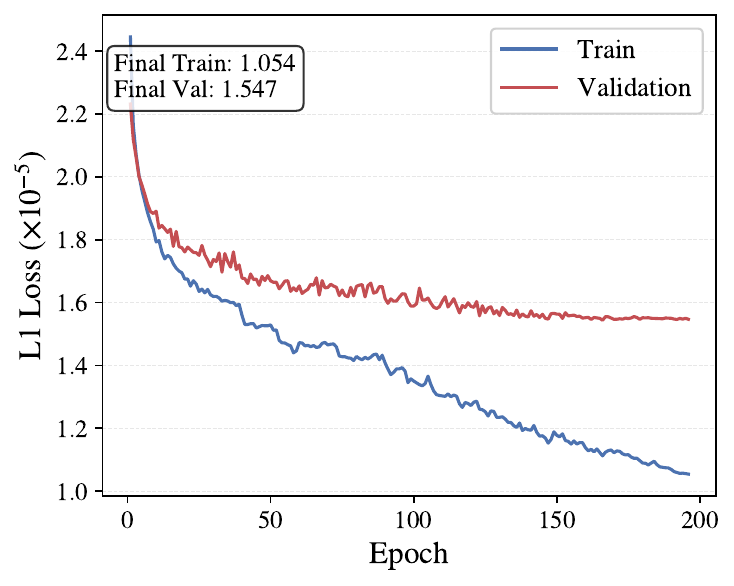}
\caption{
  Training and validation loss curves for $\rho_{N}$ (top) and $\rho_Z$ (bottom) in the CrystalFNO Solver.
}
\label{fig:fno_loss_results}
\end{figure}

Fig.~\ref{fig:fno_loss_results} presents the training and validation loss curves for both density field models, $\rho_{N}$ and $\rho_Z$. Both models exhibit stable convergence throughout training, with validation losses closely tracking the training losses without significant divergence, indicating successful optimization and no significant overfitting. The final validation losses remain at low levels (approximately $1.5 \times 10^{-5}$ for $\rho_{N}$ and $1.6 \times 10^{-5}$ for $\rho_Z$), with no significant gap between training and validation losses, confirming that the models are well-converged.

To further evaluate the model's generalization across different structure types, we group the test set by crystal system, with results summarized in Table~\ref{tab:l1_by_crystal}. The error patterns for $\rho_{N}$ and $\rho_Z$ are highly consistent across all crystal systems. The cubic system contains the largest number of samples (1,055) and exhibits the smallest prediction errors ($\rho_{N}$: $6.15 \times 10^{-6}$, $\rho_Z$: $6.46 \times 10^{-6}$); the triclinic system contains the fewest samples (217) and exhibits the largest errors ($\rho_{N}$: $2.36 \times 10^{-5}$, $\rho_Z$: $2.24 \times 10^{-5}$).

These results suggest that high-symmetry structures are easier to learn than low-symmetry structures, even when sample sizes differ substantially. High-symmetry systems, such as cubic and hexagonal crystals, impose stronger constraints on atomic arrangements and produce more regular density fields, whereas low-symmetry systems, such as monoclinic and triclinic crystals, allow greater structural variability and are more difficult to predict. The consistently higher accuracy for high-symmetry systems suggests that crystal symmetry may be more influential than dataset size in determining prediction accuracy.

\begin{table}[htbp]
\centering
\caption{Mean $L^1$ loss of the CrystalFNO Solver on the test set grouped by crystal system. Values are in units of $10^{-5}$.}
\label{tab:l1_by_crystal}
\begin{tabular}{lccc}
\toprule
Crystal System & Count & $\rho_{N}$ $L^1$ & $\rho_Z$ $L^1$ \\
\midrule
Cubic & 1055 & $0.615$ & $0.646$ \\
Hexagonal & 456  & $1.29$ & $1.32$ \\
Tetragonal & 741  & $1.34$ & $1.33$ \\
Trigonal & 411  & $1.95$ & $1.83$ \\
Orthorhombic & 875  & $1.94$ & $1.97$ \\
Monoclinic & 687  & $2.26$ & $2.20$ \\
Triclinic & 217  & $2.36$ & $2.24$ \\
\midrule
Total & 4442 & $1.53$ & $1.52$ \\
\bottomrule
\end{tabular}
\end{table}

\begin{figure*}[htbp]
\centering
\includegraphics[width=0.8\textwidth]{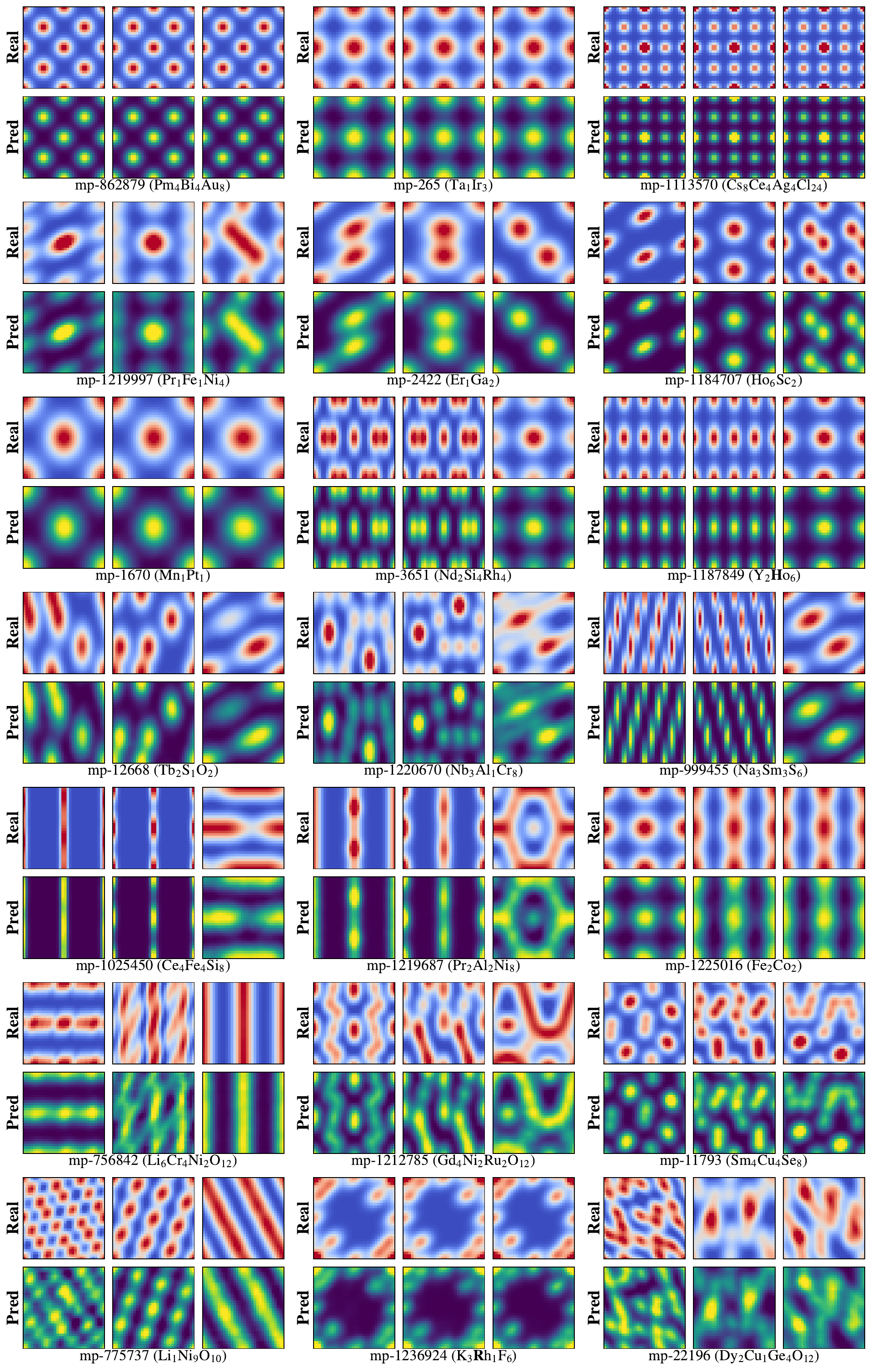}
\caption{Density-field comparison for the five best-matching structures in each crystal system obtained from the CrystalFNO Solver. Rows from top to bottom correspond to cubic, hexagonal, tetragonal, trigonal, orthorhombic, monoclinic, and triclinic systems. For each structure, the top and bottom panels show the ground-truth and CrystalFNO Solver-predicted fields, respectively, with columns corresponding to the $X$, $Y$, and $Z$ projections.}
\label{fig:best_by_crystal}
\end{figure*}

\begin{figure*}[htbp]
\centering
\includegraphics[width=0.8\textwidth]{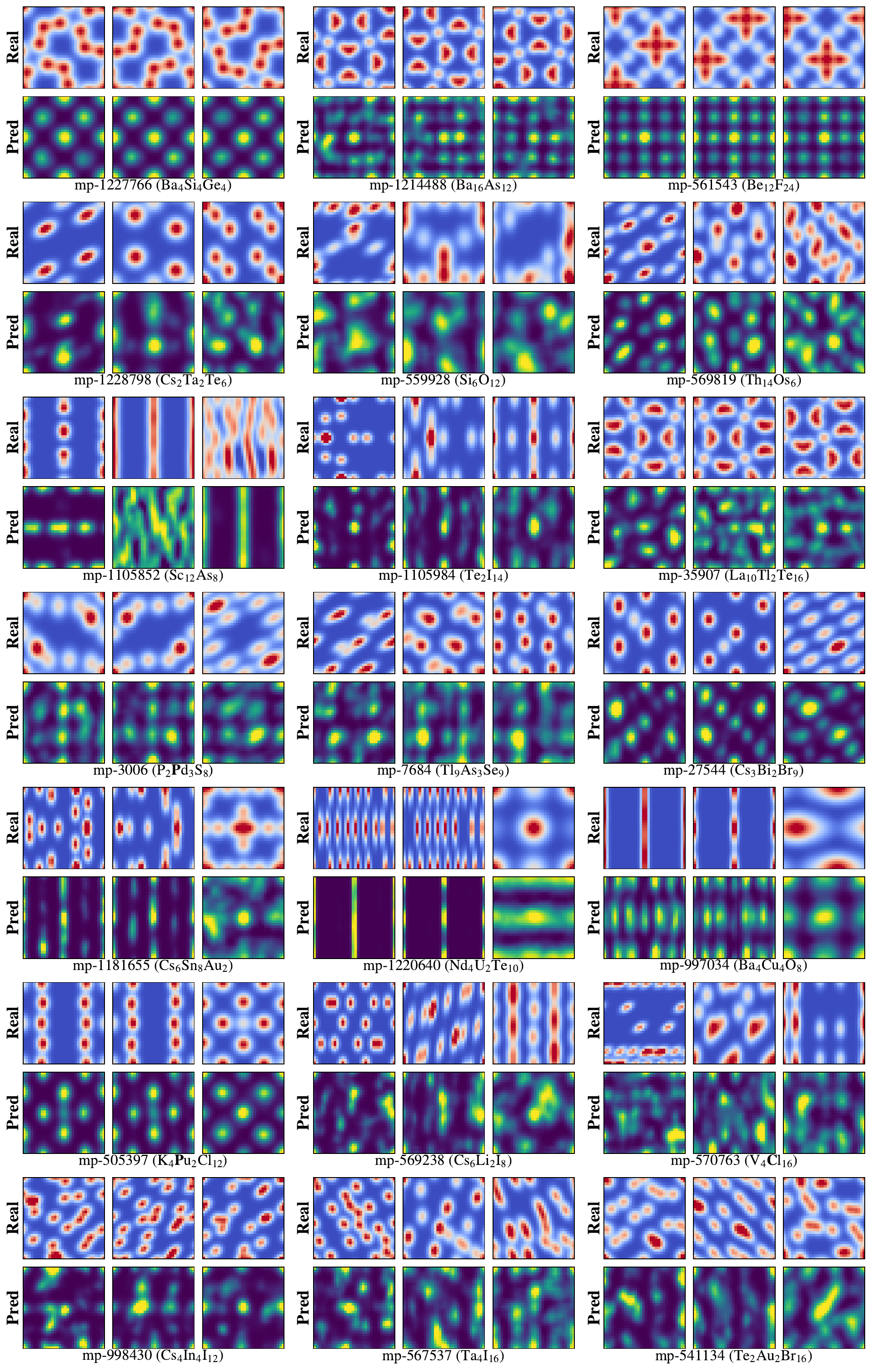}
\caption{Density-field comparison for the five worst-matching structures in each crystal system obtained from the CrystalFNO Solver. Rows from top to bottom correspond to cubic, hexagonal, tetragonal, trigonal, orthorhombic, monoclinic, and triclinic systems. For each structure, the top and bottom panels show the ground-truth and CrystalFNO Solver-predicted fields, respectively, with columns corresponding to the $X$, $Y$, and $Z$ projections.}
\label{fig:worst_by_crystal}
\end{figure*}

To visually assess the CrystalFNO Solver across crystal systems, we select the five best and five worst test structures from each system according to the $\rho_N$ $L^1$ loss. Figures~\ref{fig:best_by_crystal} and~\ref{fig:worst_by_crystal} show the selected structures across all seven crystal systems. Each structure is shown from three orthogonal projections ($X$, $Y$, and $Z$), with the ground-truth density field above the CrystalFNO Solver prediction. Within each crystal system, the five structures are ordered by increasing $L^1$ loss.

\paragraph{Best matching.} As shown in Fig.~\ref{fig:best_by_crystal}, the CrystalFNO Solver accurately reproduces the major features of the density fields across all seven crystal systems. The close correspondence between the ground-truth and predicted fields across the $X$, $Y$, and $Z$ projections indicates that the periodic boundary conditions are well represented, with no apparent spurious discontinuities at the cell boundaries for representative cubic $\mathrm{Pm_4Bi_4Au_8}$, orthorhombic $\mathrm{Ce_4Fe_4Si_8}$, and triclinic $\mathrm{LiNi_9O_{10}}$ structures.

The performance is maintained across different crystal systems and unit-cell sizes. For example, the cubic structures $\mathrm{Pm_4Bi_4Au_8}$, $\mathrm{TaIr_3}$, and $\mathrm{Cs_8Ce_4Ag_4Cl_{24}}$ contain different numbers of atoms per unit cell, yet their periodic density modulations and dominant density maxima are consistently reproduced across all three projections, with only minor differences in density amplitude. The model also resolves structurally complex, multicomponent crystals with anisotropic density distributions. For orthorhombic $\mathrm{Ce_4Fe_4Si_8}$, $\mathrm{Pr_2Al_2Ni_8}$, and $\mathrm{Fe_2Co_2}$, the elongated density maxima and paired features in the ground-truth fields are well reproduced. For lower-symmetry monoclinic $\mathrm{Li_6Cr_4Ni_2O_{12}}$ and triclinic $\mathrm{LiNi_9O_{10}}$, the global arrangement of fragmented and irregular density extrema is likewise preserved.

These examples demonstrate that the CrystalFNO Solver is not restricted to simple elemental metallic structures, but can also capture complex density topologies in chemically diverse compounds, including intermetallics, oxides, and nonmetallic systems such as $\mathrm{Na_3Sm_3S_6}$, $\mathrm{Sm_4Cu_4Se_8}$, and $\mathrm{Dy_2CuGe_4O_{12}}$. Overall, the results demonstrate robust prediction of periodic crystal density fields across diverse symmetries, unit-cell sizes, structural complexities, and chemical environments.

\paragraph{Worst matching.} As shown in Fig.~\ref{fig:worst_by_crystal}, we examine the five structures with the largest $\rho_N$ $L^1$ reconstruction losses within each crystal system. Despite these large errors, the CrystalFNO Solver generally preserves the periodicity of the density fields, with no apparent spurious discontinuities at the unit-cell boundaries for representative cubic $\mathrm{Ba_4Si_4Ge_4}$, orthorhombic $\mathrm{Cs_2Ta_2Te_6}$, and triclinic $\mathrm{Te_2Au_2Br_{16}}$ structures.

Several high-loss structures retain recognizable global density organization. For cubic $\mathrm{Ba_4Si_4Ge_4}$ and monoclinic $\mathrm{K_4Pu_2Cl_{12}}$, the predicted fields preserve clear periodic ordering and recognizable density features, although they do not closely match the ground truth. In contrast, for hexagonal $\mathrm{Si_6O_{12}}$, tetragonal $\mathrm{La_{10}Tl_2Te_{16}}$, and triclinic $\mathrm{Te_2Au_2Br_{16}}$, the predicted density fields become substantially blurred, with many atomic-scale features poorly resolved or absent. These deviations are particularly pronounced for the lower-symmetry structures and are consistent with the larger $L^1$ losses reported for triclinic systems in Table~\ref{tab:l1_by_crystal}.

Overall, the high-loss examples reveal two distinct error regimes: global lattice-scale ordering can remain well captured despite amplitude and shape errors, whereas fine-scale and irregular density topologies are more difficult to reconstruct. Thus, while the CrystalFNO Solver robustly preserves the periodic character of crystal density fields, its ability to resolve complex local density features remains limited for certain challenging structures.

\begin{figure}[htbp]
\centering
\includegraphics[width=0.5\textwidth]{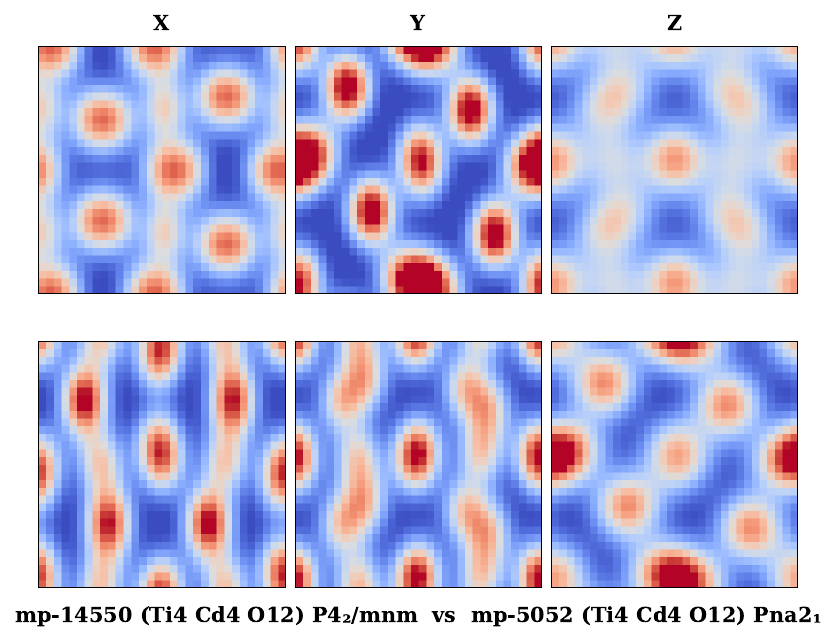}
\caption{Representative case illustrating the ill-posed nature of density-field prediction arising from polymorphism. The $\mathrm{Ti_4Cd_4O_{12}}$ structures mp-14550 and mp-5052 share the same chemical formula and have similar lattice parameters but adopt different space groups ($P4_{2}/mnm$ and $Pna2_{1}$, respectively), resulting in distinct density fields. Columns correspond to the $X$, $Y$, and $Z$ projections.}
\label{fig:illposed_polymorph}
\end{figure}

\paragraph{Error Sources.} The global density mismatches observed in Fig.~\ref{fig:worst_by_crystal} arise partly from the nonunique mapping between chemical formula, lattice parameters, and atomic configuration. Distinct atomic arrangements can share the same composition and lattice parameters but produce different density fields. Part of this ambiguity originates from global translation and discrete symmetry transformations, which can change the density field while leaving the chemical formula and lattice parameters unchanged. In this work, all structures are Niggli reduced and origin standardized, so these two sources of ambiguity are largely eliminated. The remaining error is therefore dominated by polymorphism, where the same chemical formula and similar lattice parameters can correspond to different space groups and atomic arrangements, giving rise to substantially different density fields. Figure~\ref{fig:illposed_polymorph} shows a representative example.

As shown in Fig.~\ref{fig:illposed_polymorph}, the $\mathrm{Ti_4Cd_4O_{12}}$ structures mp-14550 and mp-5052 have the same chemical formula and similar lattice parameters, but different space groups and atomic arrangements, resulting in substantially different density fields. This demonstrates that chemical formula and lattice parameters do not uniquely determine the density field, imposing a fundamental limitation on deterministic CrystalFNO Solver prediction.

Because the CrystalFNO Solver produces a single deterministic output for each input, it cannot distinguish polymorphs that share the same chemical formula and lattice parameters. Consequently, predictions for configurations that are poorly represented in the training distribution may be biased toward more frequently observed density patterns, contributing to the deviations in Fig.~\ref{fig:worst_by_crystal}. This ambiguity could be mitigated by incorporating additional structural information, such as space group, as conditional inputs to better distinguish alternative atomic configurations.

\subsection{Post-Processing Structure Reconstruction Accuracy}
To assess whether the CrystalFNO Solver-predicted density fields are sufficiently accurate for structure reconstruction, we applied the complete post-processing pipeline to the predicted $\rho_N$ and $\rho_Z$ fields. Of the 4{,}442 test structures, 4{,}294 were successfully reconstructed and evaluated; the remaining 148 failed during peak detection or position optimization. The resulting structures were evaluated using positional RMSE and element match rate across the seven crystal systems, as summarized in Table~\ref{tab:postprocess_by_crystal}.

\begin{table}[htbp]
\centering
\caption{Post-processing reconstruction accuracy by crystal system using CrystalFNO Solver-predicted density fields.}
\label{tab:postprocess_by_crystal}
\begin{tabular}{lccc}
\toprule
Crystal System & Count & RMSE (\AA) & Match Rate (\%) \\
\midrule
Cubic & 1039 & 0.247 & 90.5 \\
Hexagonal & 453 & 0.552 & 79.7 \\
Tetragonal & 733 & 0.606 & 78.6 \\
Trigonal & 391 & 0.917 & 74.7 \\
Orthorhombic & 829 & 1.023 & 67.2 \\
Monoclinic & 633 & 1.195 & 65.3 \\
Triclinic & 216 & 1.025 & 65.9 \\
\midrule
Total & 4294 & 0.730 & 76.4 \\
\bottomrule
\end{tabular}
\end{table}

The reconstruction accuracy varies substantially across crystal systems. Cubic systems exhibit the best performance (RMSE = $0.247$~\AA, match rate = $90.5\%$), whereas monoclinic systems have the largest positional error (RMSE = $1.195$~\AA, match rate ($65.3\%$). This trend is consistent with the density-field prediction errors in Table~\ref{tab:l1_by_crystal}, suggesting that errors in the predicted density fields carry over to the subsequent structure reconstruction.

To further evaluate reconstruction success under combined accuracy requirements, Table~\ref{tab:dual_threshold} reports the fraction of structures satisfying both the RMSE and element match-rate thresholds. Relaxing the RMSE threshold from $0.2$ to $1.0$~\AA\ increases the success rate by approximately $4$--$14$ percentage points depending on the match-rate threshold, whereas relaxing the match-rate requirement from $100\%$ to $60\%$ yields an increase of roughly $12$--$21$ percentage points depending on the RMSE threshold. Thus, elemental assignment imposes a more stringent constraint on successful reconstruction than positional accuracy. Even under the most lenient criteria (RMSE $\leq 1.0$~\AA, match rate $\geq 60\%$), only $53.3\%$ of structures satisfy both requirements, highlighting the difficulty of recovering complete crystal structures from predicted density fields.

\begin{table}[htbp]
\centering
\caption{Percentage of structures meeting combined RMSE and element match rate criteria on the test set. Each cell reports the fraction of materials (\%) that simultaneously satisfy RMSE $\le$ row threshold and match rate $\ge$ column threshold.}
\label{tab:dual_threshold}
\begin{tabular}{cccccc}
\toprule
\multirow{2}{*}{RMSE (\AA)} & \multicolumn{5}{c}{Element Match Rate Requirement} \\
\cmidrule(lr){2-6}
 & $\ge 60\%$ & $\ge 70\%$ & $\ge 80\%$ & $\ge 90\%$ & $\ge 100\%$ \\
\midrule
0.2 & 39.8\% & 37.1\% & 34.4\% & 29.6\% & 28.1\% \\
0.4 & 43.8\% & 40.4\% & 37.3\% & 32.0\% & 30.1\% \\
0.6 & 45.9\% & 42.1\% & 38.8\% & 32.9\% & 30.8\% \\
0.8 & 48.9\% & 44.5\% & 40.3\% & 33.9\% & 31.6\% \\
1.0 & 53.3\% & 47.8\% & 42.4\% & 35.2\% & 32.4\% \\
\bottomrule
\end{tabular}
\end{table}

Overall, these results show that the CrystalFNO Solver-predicted density fields provide a viable basis for crystal structure reconstruction, while post-processing substantially improves the recovered structures. The remaining performance gap across crystal systems and the sensitivity to elemental assignment highlight key directions for further improvement.

\begin{figure*}[htbp]
\centering
\includegraphics[width=0.95\textwidth]{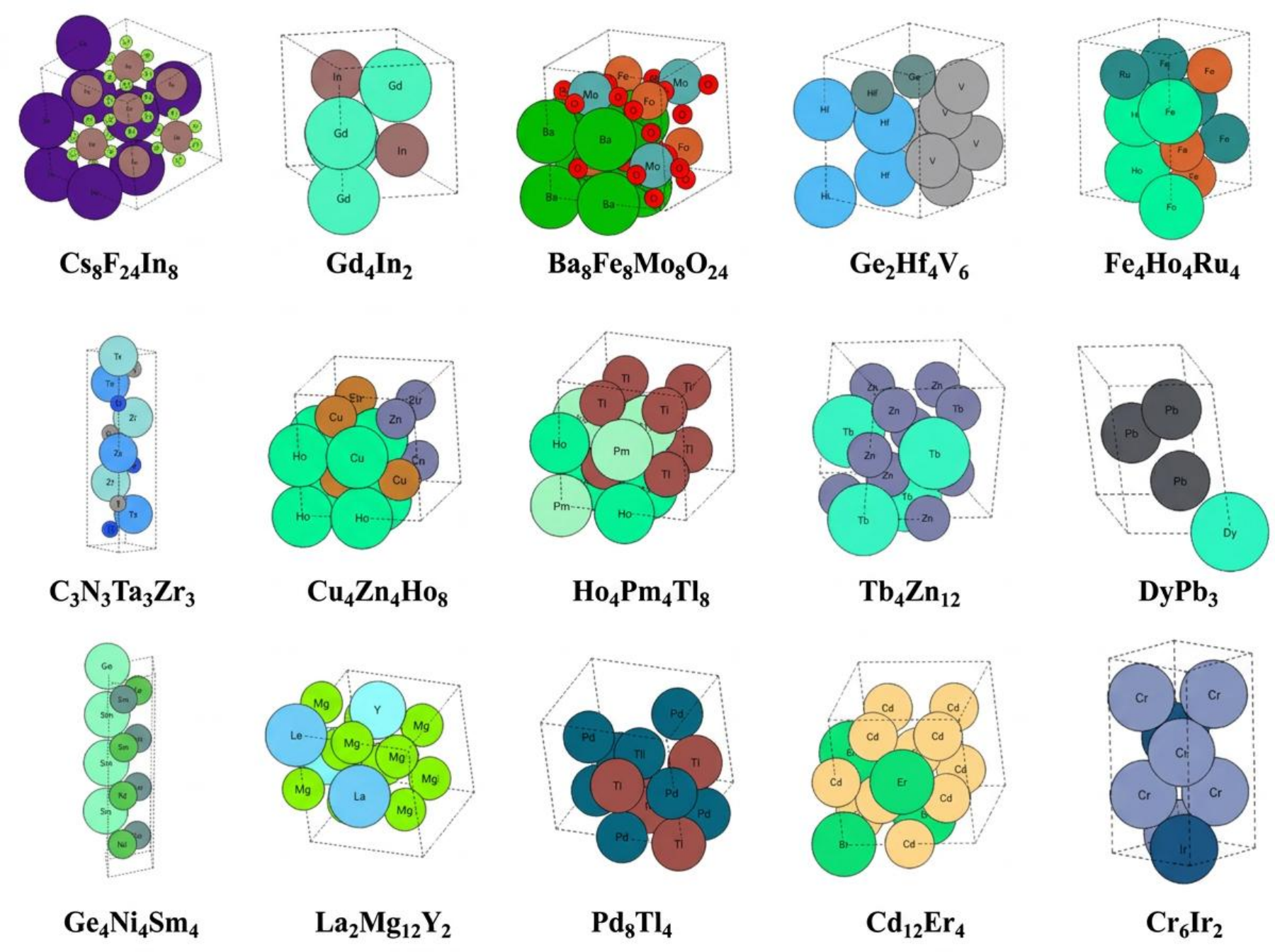}
\caption{Selected novel crystal structures generated by the CrystalFNO Solver that pass the DPA relaxation screening. Each structure is displayed with its unit cell, with atoms color-coded by element type. These structures span a diverse range of crystal systems and chemical compositions, demonstrating the ability of the CrystalFNO Solver to generate geometrically reasonable and energetically favorable crystal structures.}
\label{fig:generated_structures}
\end{figure*}

\subsection{End-to-End Crystal Structure Generation and Screening}
\label{sec:3c}
The complete workflow was applied to the 4{,}380 unique chemical formulas in the test set. Multi-batch sampling generated 43{,}686 candidate structures, of which 37{,}049 converged during DPA3 relaxation \cite{Zhang2018DP,Wang2018DP}. Applying the structural validation criteria yielded 3{,}663 structures, and subsequent deduplication identified 320 structures not present in the Materials Project database \cite{Jain2013MP}, spanning 104 distinct chemical formulas. Representative examples include $\mathrm{AgLaSi}$ and $\mathrm{Ba_4Nb_2O_{12}V_2}$.

Figure~\ref{fig:generated_structures} shows representative generated structures that passed the screening. The candidates span diverse crystal systems and chemical compositions, demonstrating the ability of the workflow to produce structurally consistent crystal candidates from chemical formulas.

DPA3 single-point energies were obtained for all 320 candidates without failed calculations. The uncorrected formation energies range from $-2.588$ to $0.170$~eV atom$^{-1}$, with a mean of $-0.227$~eV atom$^{-1}$ and a median of $-0.060$~eV atom$^{-1}$. Of the 320 candidates, 179 ($55.9\%$) have negative formation energies relative to the selected elemental reference states. Because these energies are evaluated at fixed geometries, negative formation energies indicate favorable energetics relative to the constituent elements but do not establish thermodynamic stability against competing phases.

\subsection{Generalization to unseen chemical formulas}
\label{sec:unseen}

To test whether the framework generalizes beyond the training distribution, we applied the complete pipeline to 3{,}000 chemical formulas that do not appear in the MP20 training, validation, or test splits. The candidate formulas were generated by an independent composition-enumeration pipeline (Appendix~\ref{app:candidate_formulas}): charge-balanced stoichiometries over three composition spaces --- oxychalcogenides (1{,}000), phosphorus chalcogenides (1{,}000), and double perovskites (1{,}000) --- screened for chemical plausibility with SMACT~\cite{smact_joss,smact_chem} and filtered against the union of the Crystallography Open Database~\cite{cod}, the Materials Project~\cite{Jain2013MP}, and the Open Quantum Materials Database~\cite{oqmd} (3{,}597{,}064 unique known formulas). None of the 3{,}000 formulas therefore appears in any of the three reference databases at the level of the reduced chemical formula, and none appears in the MP20 splits used to train or evaluate CrystalFNO.

For each formula, the CVAE generates 100 candidate lattices. CrystalFNO predicts the corresponding number-density and electron-density fields for each lattice, and the post-processing pipeline reconstructs atomic positions and element types. The reconstructed structures are then ranked by their voxel-level $L^{1}$ loss against the CrystalFNO-predicted density fields, and the 10 candidates with the lowest loss are retained for DPA3 relaxation and subsequent screening, following the same procedure described in Sec.~\ref{sec:3c}.

\paragraph{Generation statistics.} Across the 3{,}000 formulas, the pipeline produced 29{,}910 valid crystal structures. After DPA3 relaxation, 24{,}582 structures converged successfully, of which 20{,}150 reached convergence within 200 relaxation steps. Formation energies were computed for all relaxed structures, and 18{,}363 of them exhibit a formation energy below $-0.2$~eV~atom$^{-1}$. Table~\ref{tab:unseen_summary} summarizes these results.

\begin{table}[htbp]
\centering
\caption{Overall generation statistics on the 3{,}000 unseen chemical formulas.}
\label{tab:unseen_summary}
\begin{tabular}{lc}
\toprule
Quantity & Value \\
\midrule
Chemical formulas tested                        & 3{,}000 \\
Valid crystal structures generated              & 29{,}910 \\
Structures converged after DPA3 relaxation      & 24{,}582 \\
\quad of which converged within 200 steps       & 20{,}150 \\
Structures with $E_f < -0.2$~eV~atom$^{-1}$     & 18{,}363 \\
Structures with $E_f < 0.0$~eV~atom$^{-1}$      & 22{,}518 \\
\bottomrule
\end{tabular}
\end{table}

\paragraph{Structure validity.} The DPA3 relaxation was performed with the same settings as in Appendix~\ref{app:dpa}. Of the 29{,}910 generated structures, 24{,}582 converged successfully, corresponding to a convergence rate of $82.2\%$. The remaining structures either did not reach the force threshold within the maximum number of steps or failed the geometric validation criteria (lattice change $<5\%$, atomic RMSD $<0.3$~\AA). Among the converged structures, 20{,}150 ($82.0\%$) reached convergence within 200 steps, indicating that most reconstructed structures are already close to a local energy minimum before relaxation.

\paragraph{Energetics.} Formation energies were computed with DPA3 for all relaxed structures. Of the 24{,}582 converged structures, 18{,}363 ($74.7\%$) exhibit a formation energy below $-0.2$~eV~atom$^{-1}$ relative to the elemental reference states, and 22{,}518 ($91.6\%$) have $E_f < 0$. As in Sec.~\ref{sec:3c}, these energies are evaluated at fixed geometries and therefore indicate favorable energetics relative to the constituent elements, but do not by themselves establish thermodynamic stability against competing phases. The high fraction of structures with strongly negative formation energies nevertheless shows that the framework consistently produces energetically plausible candidates for compositions it has never encountered.

\begin{figure}[htbp]
\centering
\includegraphics[width=0.45\textwidth]{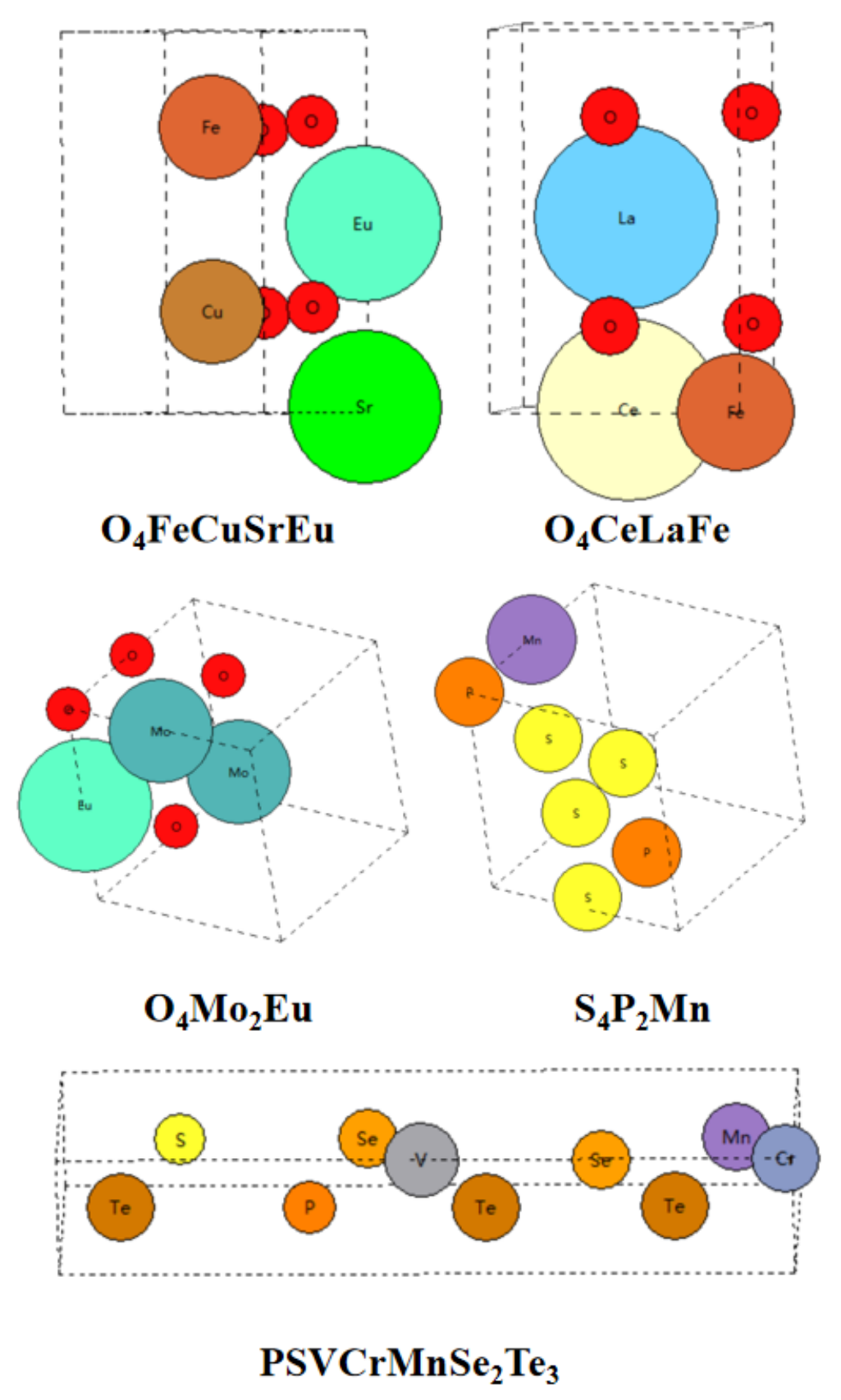}
\caption{Representative crystal structures generated for unseen chemical formulas
spanning oxychalcogenides, phosphorus chalcogenides, and double perovskites. Each
structure passed the full screening pipeline.}
\label{fig:unseen_structures}
\end{figure}

\paragraph{Representative structures.} Figure~\ref{fig:unseen_structures} shows representative generated structures for unseen formulas, spanning all three composition spaces. The candidates exhibit diverse coordination environments and reasonable bond lengths, demonstrating that the framework can generate structurally consistent crystal candidates for compositions outside the training distribution.

\section{conclusion}

In this work, we have developed CrystalFNO, a coupled generation-solving framework for composition-driven crystal structure discovery that integrates Fourier Neural Operators (FNOs) with conditional variational autoencoders (CVAEs). The framework addresses the challenge of generating periodic crystal structures from prescribed chemical compositions by separating lattice generation from atomic structure reconstruction. Specifically, the CVAE generates diverse candidate lattice parameters in a low-dimensional symmetric basis-coefficient space, while the CrystalFNO Solver predicts the corresponding number-density and electron-density fields on periodic voxel grids. These fields are subsequently converted into explicit atomic structures through peak detection, position optimization, and elemental weight optimization, followed by structural screening and machine-learning interatomic-potential relaxation. By incorporating Fourier-domain operations, the proposed solver captures global correlations in periodic crystal fields while retaining local spatial information.

We first evaluated CrystalFNO on the MP20 dataset, where the predicted density fields enabled the reconstruction of 4,294 out of 4,442 test structures. The reconstructed structures achieved an overall positional RMSE of 0.730 \AA and an element match rate of 76.4\%, demonstrating the feasibility of recovering atomic configurations from predicted density fields. Applying the complete generation pipeline to the test-set compositions yielded 320 distinct structures absent from the Materials Project, spanning 104 chemical formulas. More importantly, we investigated the generalization capability of CrystalFNO using 3,000 previously unseen chemical formulas independently constructed across oxychalcogenide, phosphorus chalcogenide, and double-perovskite composition spaces. The framework generated 29,910 valid crystal structures, of which 24,582 converged successfully under DPA3 relaxation. Among the converged structures, 18,363 exhibited formation energies below -0.2 eV atom\(^{-1}\) relative to the selected elemental reference states. These results demonstrate the potential of the proposed framework to generate structurally consistent crystal candidates beyond the chemical compositions represented in its training dataset.

Despite these promising results, several limitations remain. First, the reconstruction accuracy varies across crystal systems, with lower accuracy observed for structures with lower crystallographic symmetry. Second, the mapping from chemical composition and lattice parameters to density fields is inherently non-unique, particularly for compositions exhibiting polymorphism. Incorporating additional crystallographic information, such as space-group symmetry, may help distinguish alternative atomic arrangements. Furthermore, the formation energies reported here are referenced to elemental states and do not establish thermodynamic stability against competing phases. Future work will therefore focus on improving reconstruction accuracy and structural diversity, incorporating explicit symmetry constraints and additional structural descriptors, and validating promising candidates through more accurate first-principles calculations and phase-stability analysis.

Overall, CrystalFNO provides a complementary approach to existing generative models for inorganic crystals by combining composition-conditioned lattice generation with Fourier-based periodic density-field prediction and explicit atomic reconstruction. Its ability to generate candidate structures for thousands of previously unseen chemical formulas highlights the potential of operator-learning approaches for accelerating computational materials discovery.

\section*{Code Availability}
The code for CrystalFNO is openly available at \url{https://github.com/pqr-yzj/CrystalFNO}. The repository provides the complete pipeline, including data preprocessing, training of the FNO-based number-density and electron-density solvers, training of the CVAE lattice generator, structure reconstruction via peak detection and gradient-based position/weight optimization, and end-to-end CIF generation from chemical formulas.

\section*{Acknowledgements}
The authors acknowledge the use of DeepSeek and ChatGPT for coding assistance, language editing, and manuscript preparation. All scientific analyses, interpretations, methodology, and conclusions were developed independently by the authors.
\bibliography{references}

\appendix

\section{Symmetric Basis Decomposition for Lattice Parameters}
\label{app:basis_comp}

We parameterize the symmetric metric tensor using its matrix square root. We introduce six basis matrices that span the space of $3\times3$ symmetric matrices:
\begin{align}
\mathbf{B}_1 &= \begin{pmatrix}
1 & 0 & 0 \\
0 & 1 & 0 \\
0 & 0 & 1
\end{pmatrix}, &
\mathbf{B}_2 &= \begin{pmatrix}
1 & 0 & 0 \\
0 & -1 & 0 \\
0 & 0 & 0
\end{pmatrix}, &
\mathbf{B}_3 &= \begin{pmatrix}
1 & 0 & 0 \\
0 & 1 & 0 \\
0 & 0 & -2
\end{pmatrix}, \\[6pt]
\mathbf{B}_4 &= \begin{pmatrix}
0 & 1 & 0 \\
1 & 0 & 0 \\
0 & 0 & 0
\end{pmatrix}, &
\mathbf{B}_5 &= \begin{pmatrix}
0 & 0 & 1 \\
0 & 0 & 0 \\
1 & 0 & 0
\end{pmatrix}, &
\mathbf{B}_6 &= \begin{pmatrix}
0 & 0 & 0 \\
0 & 0 & 1 \\
0 & 1 & 0
\end{pmatrix}.
\end{align}
For a given lattice matrix $\bm{L}$, we compute $S = \sqrt{\bm{L}^{\top}\bm{L}}$ and project $S$ onto the above basis, yielding a unique decomposition:
\begin{equation}
S = k_1 \mathbf{B}_1 + k_2 \mathbf{B}_2 + k_3 \mathbf{B}_3
  + k_4 \mathbf{B}_4 + k_5 \mathbf{B}_5 + k_6 \mathbf{B}_6,
\end{equation}
where $\{k_i\}_{i=1}^{6}$ are the projection coefficients. The coefficient vector $\mathbf{k} = [k_1, k_2, \dots, k_6]$ serves as the six-dimensional representation of the lattice parameters.

\begin{table*}[htbp]
\centering
\caption{Unit cell characteristics and basis coefficient constraints for the seven
crystal systems.}
\label{tab:crystal_systems}
\begin{tabular}{@{}lccc@{}}
\toprule
\textbf{Crystal system} & \textbf{Space group numbers} & \textbf{Unit cell constraints} & \textbf{Coefficient constraints} \\
\midrule
Triclinic & 1--2 & $a \neq b \neq c,\ \alpha \neq \beta \neq \gamma$ & None \\
Monoclinic & 3--15 & $a \neq b \neq c,\ \alpha = \gamma = 90^\circ,\ \beta \neq 90^\circ$ & $k_4 = k_6 = 0$ \\
Orthorhombic & 16--74 & $a \neq b \neq c,\ \alpha = \beta = \gamma = 90^\circ$ & $k_4 = k_5 = k_6 = 0$ \\
Tetragonal & 75--142 & $a = b \neq c,\ \alpha = \beta = \gamma = 90^\circ$ & $k_2 = k_4 = k_5 = k_6 = 0$ \\
Rhombohedral & 143--167 & $a = b = c,\ \alpha = \beta = \gamma \neq 90^\circ$ & $k_2 = k_3 = 0,\ k_4 = k_5 = k_6$ \\
Hexagonal & 168--194 & $a = b \neq c,\ \alpha = \beta = 90^\circ,\ \gamma = 120^\circ$ & $k_1 / k_4 = -2 - \sqrt{3},\ k_2 = k_5 = k_6 = 0$ \\
Cubic & 195--230 & $a = b = c,\ \alpha = \beta = \gamma = 90^\circ$ & $k_2 = k_3 = k_4 = k_5 = k_6 = 0$ \\
\bottomrule
\end{tabular}
\end{table*}

Given a chemical formula, we can generate $\mathbf{k}$ and convert it to the corresponding lattice parameters $L$ via the inverse decomposition.

\section{Architecture of Lattice Parameter Generator}
\label{app:cvae}

The CVAE architecture is illustrated in Fig.~\ref{fig:overall_pipeline}(b). The encoder takes the concatenated basis coefficients $\bm{k}$ and chemical-condition embedding $\bm{F}$ and outputs the latent mean $\bm{\mu}_z$ and log-variance $\log\bm{\sigma}_z^2$. The latent variable is sampled using the reparameterization trick,
\begin{equation}
\bm{z}={\mu}_z+\bm{\sigma}_z\odot\bm{\epsilon},
\qquad
\bm{\epsilon}\sim\mathcal{N}(\bm{0},\bm{I}).
\end{equation}
The decoder concatenates $\bm{z}$ with $\bm{F}$ to reconstruct the basis coefficients
$\widehat{\bm{k}}$.

The CVAE is trained using a reconstruction loss and a KL-divergence regularization term,
\begin{equation}
\mathcal{L}_{\mathrm{CVAE}}=\mathcal{L}_{\mathrm{rec}}+\beta\mathcal{L}_{\mathrm{KL}},
\end{equation}
where $\mathcal{L}_{\mathrm{rec}}$ is the mean-squared error between $\bm{k}$ and $\widehat{\bm{k}}$, and $\beta$ controls the strength of the KL regularization.

Because the lattice representation is only six-dimensional, the reconstruction task does not dominate the objective, allowing the KL term to effectively regularize the latent space. The resulting stochastic latent representation enables multiple lattice parameter sets to be sampled for a given chemical formula, thereby addressing the one-to-many nature of composition--structure relationships, including polymorphism. These candidate lattices are subsequently coupled with the CrystalFNO Solver and post-processing module to generate diverse crystal structures.

To evaluate the CVAE generator's capability in lattice parameter generation, the CVAE is conditioned on the chemical formula and generates the corresponding six basis coefficients $\bm{k} = [k_1, \dots, k_6]$, from which the lattice parameters $(a, b, c, \alpha, \beta, \gamma)$ are reconstructed. On the test set, for each unique chemical formula, we generate 100 candidate lattice parameter sets and compare their overall distribution with the ground-truth structures from the Materials Project database~\cite{Jain2013MP}.

\begin{figure}[htbp]
\centering
\includegraphics[width=0.5\textwidth]{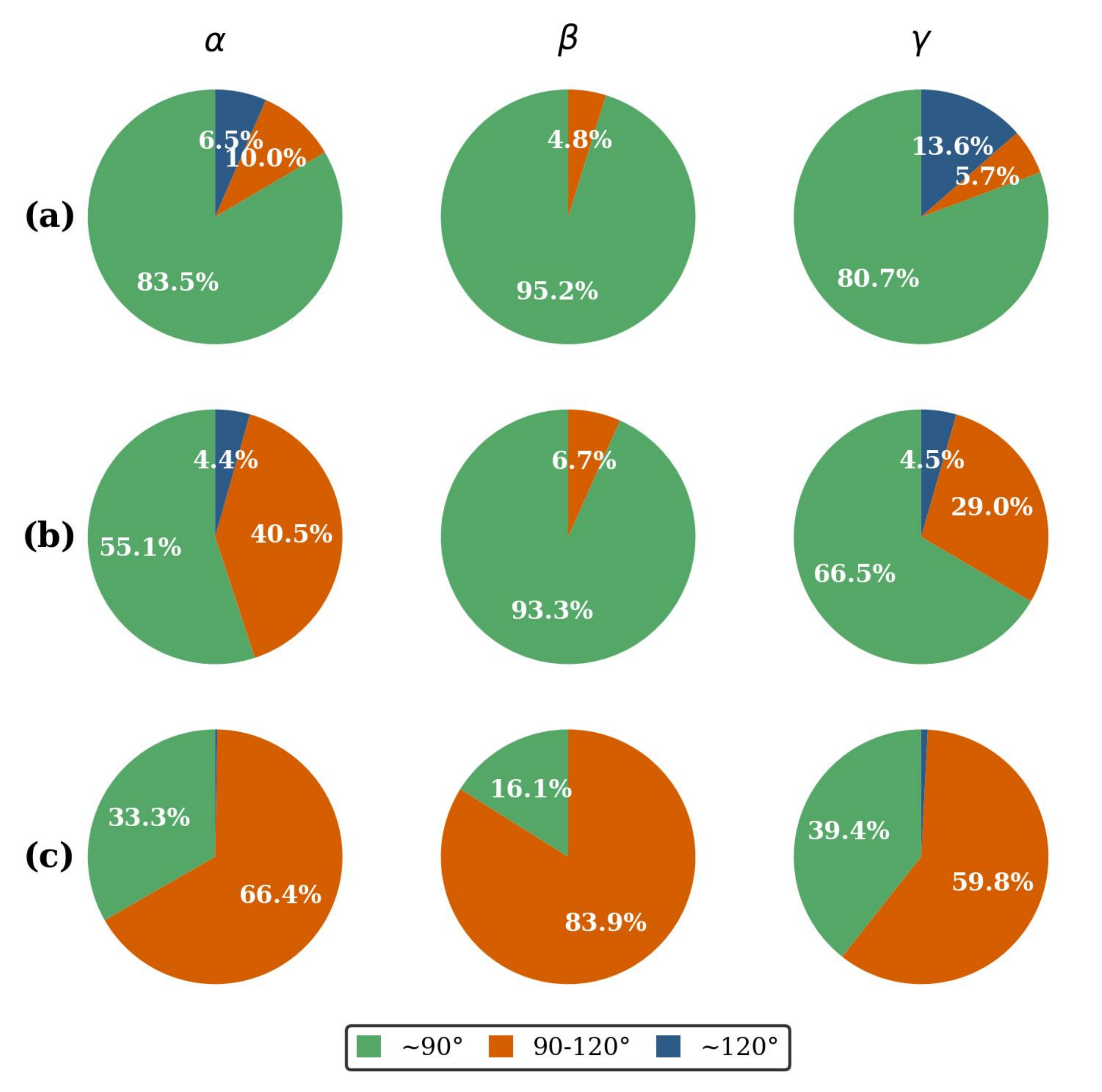}
\caption{Distribution of lattice angles \(\alpha\), \(\beta\), and \(\gamma\) across
three categories: near 90° (89.95-90.05°), 90-120° (90.05-119.9°), and near 120°
(119.9-120.1°). Each row corresponds to (a) MP ground truth, (b) symmetry-basis encoding
(ours), and (c) direct generation without basis decomposition (baseline). The
symmetry-basis encoding method achieves distributions that closely match the ground
truth, while the baseline exhibits noticeable discrepancies, particularly in the
near-90° and near-120° bins.}
\label{fig:angle_pie}
\end{figure}

Beyond the lattice lengths and volume, we further examine the angular distributions. Fig.~\ref{fig:angle_pie} presents the distribution of the three lattice angles $\alpha$, $\beta$, and $\gamma$ across three categories: near 90° (89.95--90.05°), 90--120° (90.05--119.9°), and near 120° (119.9--120.1°), visualized as pie charts. Each row corresponds to (a) the MP ground-truth distribution, (b) our symmetry-basis encoding method, and (c) the baseline method that directly generates the six raw lattice parameters without basis decomposition.

As shown in the figure, the symmetry-basis encoding method achieves distributions that closely match the MP ground truth, while the baseline exhibits noticeable discrepancies across all angles. In particular, the baseline significantly deviates from the ground truth in the near-90° and near-120° bins, underestimating or overestimating the proportions of structures with these characteristic angles. In contrast, our symmetry-basis encoding captures these discrete features with substantially higher fidelity, demonstrating that incorporating crystallographic symmetry priors through basis decomposition effectively improves the generative accuracy of lattice angles.

These results demonstrate that the CVAE generator, equipped with symmetry-basis encoding, can generate lattice parameters conditioned on the chemical formula with statistical distributions close to real data---both in terms of cell dimensions and angular distributions---providing reasonable candidate inputs for the subsequent CrystalFNO Solver density prediction and structure reconstruction.

\section{Training Details}

\begin{figure*}[htbp]
\centering
\includegraphics[width=0.7\textwidth]{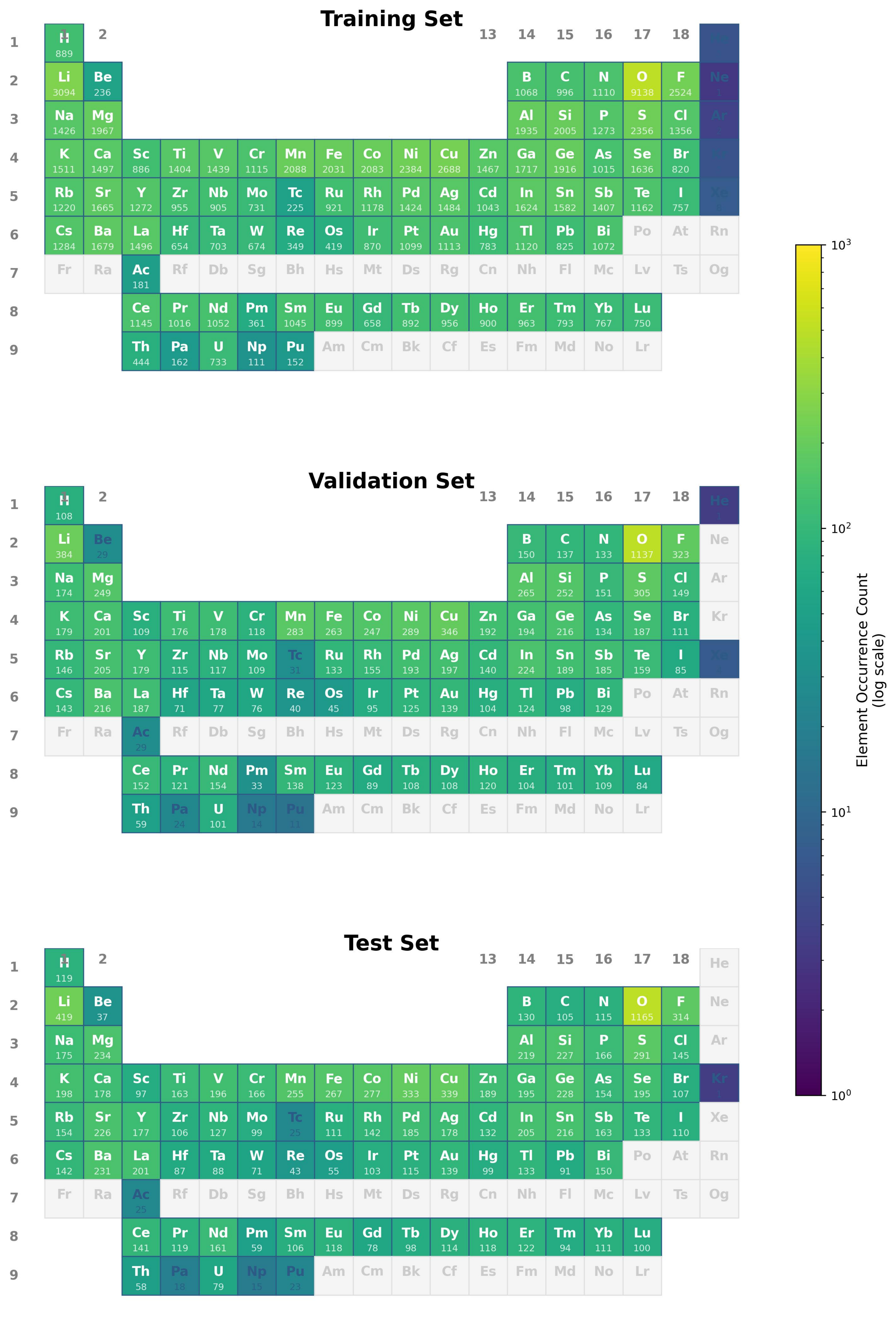}
\caption{Element distribution across training, validation, and test sets. Each cell
represents a chemical element, with color intensity indicating the number of structures
containing that element. Darker colors indicate higher occurrence frequencies. The
dataset covers a broad range of elements, with transition metals and common main-group
elements being most prevalent.}
\label{fig:element_periodic_table}
\end{figure*}

Fig.~\ref{fig:element_periodic_table} shows the element distribution across the three dataset splits. The three splits exhibit highly consistent element coverage, confirming that the data partitioning preserves the overall chemical diversity of the full dataset. The most frequently occurring elements include transition metals (Fe, Ti, Ni, Cu, Zn) and common main-group elements (O, C, N, Si), consistent with the composition distribution of the Materials Project database. The consistent distribution across splits ensures that the model is trained and evaluated on chemically representative subsets, facilitating fair assessment of generalization performance.

The CrystalFNO Solver-based density solver consists of two independently trained models: one for \(\rho_{N}\) and one for $\rho_Z$. Both models are trained using the Adam optimizer with an initial learning rate of \(8 \times 10^{-4}\), a batch size of 120, and a weight decay of \(1 \times 10^{-4}\). The learning rate is reduced by a factor of 0.8 when the validation loss plateaus for 5 epochs. All models are trained on a single NVIDIA GPU for up to 500 epochs with an early stopping patience of 30 epochs.

The CVAE generator for lattice parameter generation uses a latent dimension of 16 and a conditional embedding dimension of 32. It is trained using the AdamW optimizer with a learning rate of \(1 \times 10^{-3}\), a batch size of 128, and a weight decay of \(1 \times 10^{-5}\). To address the imbalance between orthogonal and non-orthogonal crystal systems, we apply a weighted sampling strategy that oversamples non-orthogonal structures by a factor of 30. The model is trained for up to 10,000 epochs with an early stopping patience of 50 epochs on the validation loss.

\section{DPA Details}
\label{app:dpa}

All crystal structures were geometrically optimized at zero temperature under periodic boundary conditions using the pretrained DPA-3.1-MPtrj machine-learning interatomic potential \cite{Zhang2018DP,Wang2018DP,DPA3}. The model was implemented through DeePMD-kit 3.1.2 \cite{Wang2018DP}, and structural optimization was performed using the Atomic Simulation Environment (ASE) 3.26.0 \cite{hjorth2017atomic}. Both atomic coordinates and the six independent cell-strain components were optimized using the ExpCellFilter at zero external pressure, without symmetry constraints. Optimization was performed with the BFGS algorithm until the ASE-reported maximum force, $f_{\max}$, fell below $5\times10^{-3}$~eV\,$\mathrm{\AA}^{-1}$, with a maximum of 500 steps.

For formation-energy calculations, the same DPA model and energy-evaluation procedure were used to obtain the static total potential energy $E_{\mathrm{DPA}}$ of each structure at its fixed input geometry. The formation energy per atom was calculated as
\begin{equation}
\Delta E_{\mathrm{f}}=\frac{E_{\mathrm{DPA}}-\sum_i n_i\mu_i}{N},
\end{equation}
where $n_i$ is the number of atoms of element $i$, $N$ is the total number of atoms, and $\mu_i$ is the corresponding elemental reference energy per atom. The elemental reference energies were taken from the Materials Project v2022.10.28 data snapshot \cite{Jain2013MP} without additional empirical corrections. We refer to the resulting quantity as the uncorrected DPA3 formation energy referenced to Materials Project elemental energies. A negative value indicates favorable energetics relative to the selected separated elemental reference states, but does not establish thermodynamic stability against competing compounds or phases.

\section{Candidate chemical-formula generation}
\label{app:candidate_formulas}

This appendix describes how the 3{,}000 unseen chemical formulas used in Section~\ref{sec:unseen} were constructed.

\paragraph{Composition spaces.} Three composition spaces were defined in a configuration file: oxychalcogenides, phosphorus chalcogenides, and double perovskites. Each space specifies a cation pool, an anion pool, per-element oxidation states used for charge balancing, and enumeration constraints.

\begin{itemize}[leftmargin=*,itemsep=2pt,topsep=2pt]
  \item \textbf{Oxychalcogenides}: cations La, Ce, Pr, Nd, Sm, Eu, Gd, Ti, V, Cr, Mn, Fe, Co, Ni (14); anions O, S, Se, Te with O required; max stoichiometry 8, max atoms 20, max cation species 3.
  \item \textbf{Phosphorus chalcogenides}: cations Ti, V, Cr, Mn, Fe, Co, Ni, Cu (8); anions P, S, Se, Te with P required; max stoichiometry 8, max atoms 20, max cation species 4.
  \item \textbf{Double perovskites}: cations La, Nd, Sm, Eu, Gd, Ca, Sr, Ba, Ti, V, Cr, Mn, Fe, Co, Ni, Cu, Nb, Mo, W (19); anion O; max stoichiometry 6, max atoms 10, max cation species 4.
\end{itemize}

\paragraph{Generation pipeline.} For each composition space, the pipeline has four stages:

\begin{enumerate}[label=(\arabic*)]
  \item \textbf{Enumerate} charge-balanced integer stoichiometries over the cation and
        anion subsets, with a per-combination cap on the number of formulas to prevent
        any single element combination from dominating the pool.
  \item \textbf{Screen} every candidate with SMACT~4.0.0~\cite{smact_joss,smact_chem},
        which checks charge neutrality under the ICSD-2024 oxidation-state set and
        the Pauling electronegativity rule.
  \item \textbf{Filter} out any formula whose canonical reduced form already appears
        in the Crystallography Open Database~\cite{cod}, the Materials
        Project~\cite{Jain2013MP}, or the Open Quantum Materials
        Database~\cite{oqmd}. The union of these three databases contains 3{,}597{,}064
        unique reduced formulas.
  \item \textbf{Select} a maximally spread-out subset of 1{,}000 formulas per space
        using greedy farthest-point sampling~\cite{gonzalez} over a combined
        stoichiometric and element-identity distance based on the Jaccard
        index~\cite{jaccard}.
\end{enumerate}

Table~\ref{tab:runs_app} summarizes the statistics of the three runs.

\begin{table}[htbp]
\centering
\scriptsize
\begin{tabular}{@{}l r r r r@{}}
\toprule
System & Pool & Known & Novel & Delivered \\
\midrule
Oxychalcogenides         & 14{,}975 &   830 & 14{,}145 & 1{,}000 \\
Phosphorus chalcogenides &  6{,}768 &    81 &  6{,}687 & 1{,}000 \\
Double perovskites       & 11{,}019 & 2{,}088 &  8{,}931 & 1{,}000 \\
\bottomrule
\end{tabular}
\caption{Statistics of the three novelty-filtered runs. The ``Known'' column gives the number of formulas already present in COD/MP/OQMD; ``Novel'' gives the remainder; ``Delivered'' is the number selected by farthest-point sampling.}
\label{tab:runs_app}
\end{table}

\paragraph{Scope and limitations.} ``Novel'' here means \emph{unseen in COD, MP, and OQMD at the level of the reduced chemical formula}. It is not a claim of thermodynamic stability, synthesizability, or that the material has never been made. Charge neutrality and the Pauling electronegativity rule are necessary but not sufficient conditions for a real compound. Two polymorphs of the same composition are counted as a single known formula.

\section{Additional Density Field Prediction Results}

\begin{figure*}[htbp]
\centering
\includegraphics[width=0.9\textwidth]{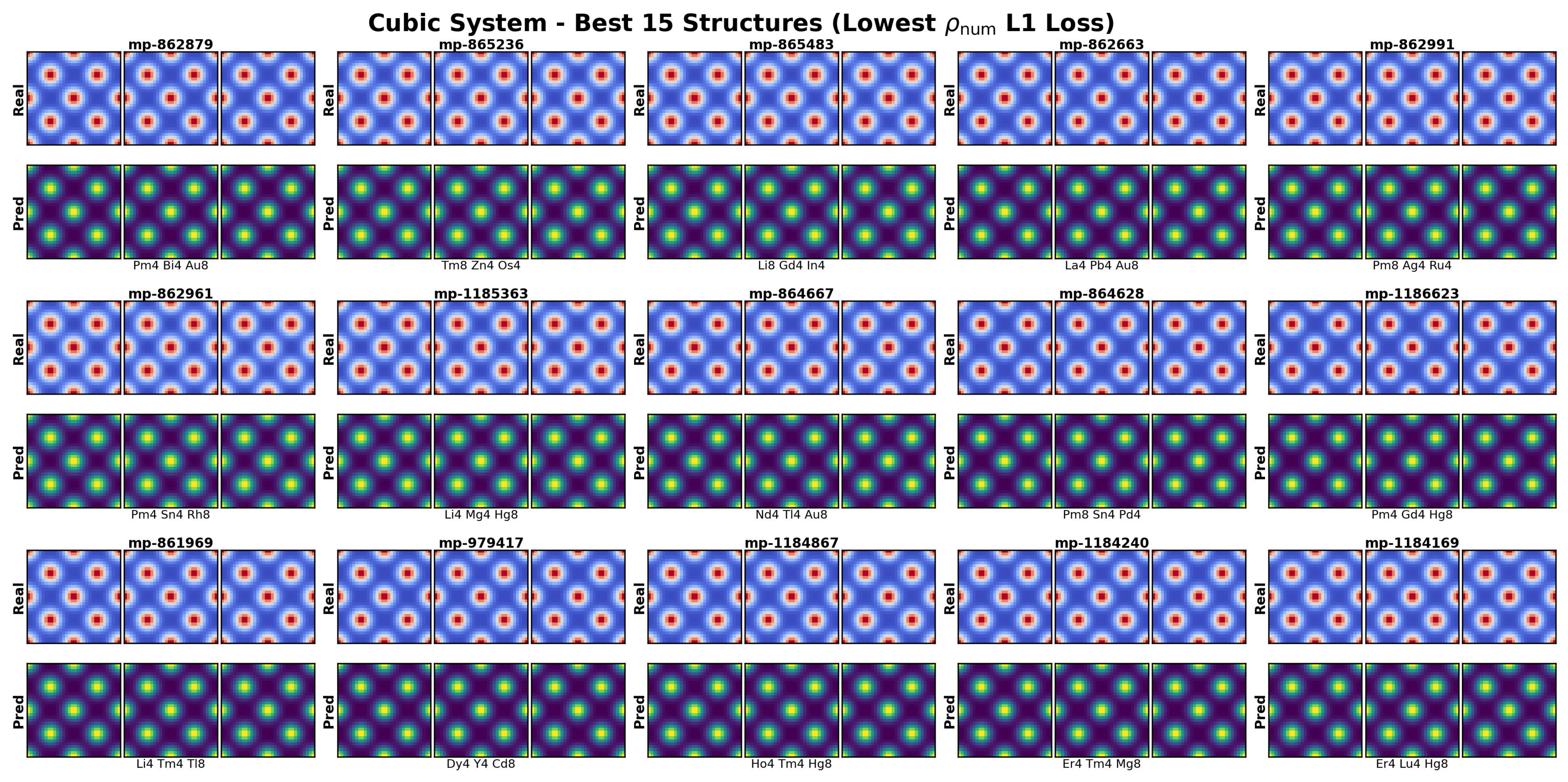}
\caption{Additional density field comparison for 15 representative cubic structures.
Top: ground truth; bottom: CrystalFNO Solver predictions.}
\label{fig:app_cubic}
\end{figure*}

\begin{figure*}[htbp]
\centering
\includegraphics[width=0.9\textwidth]{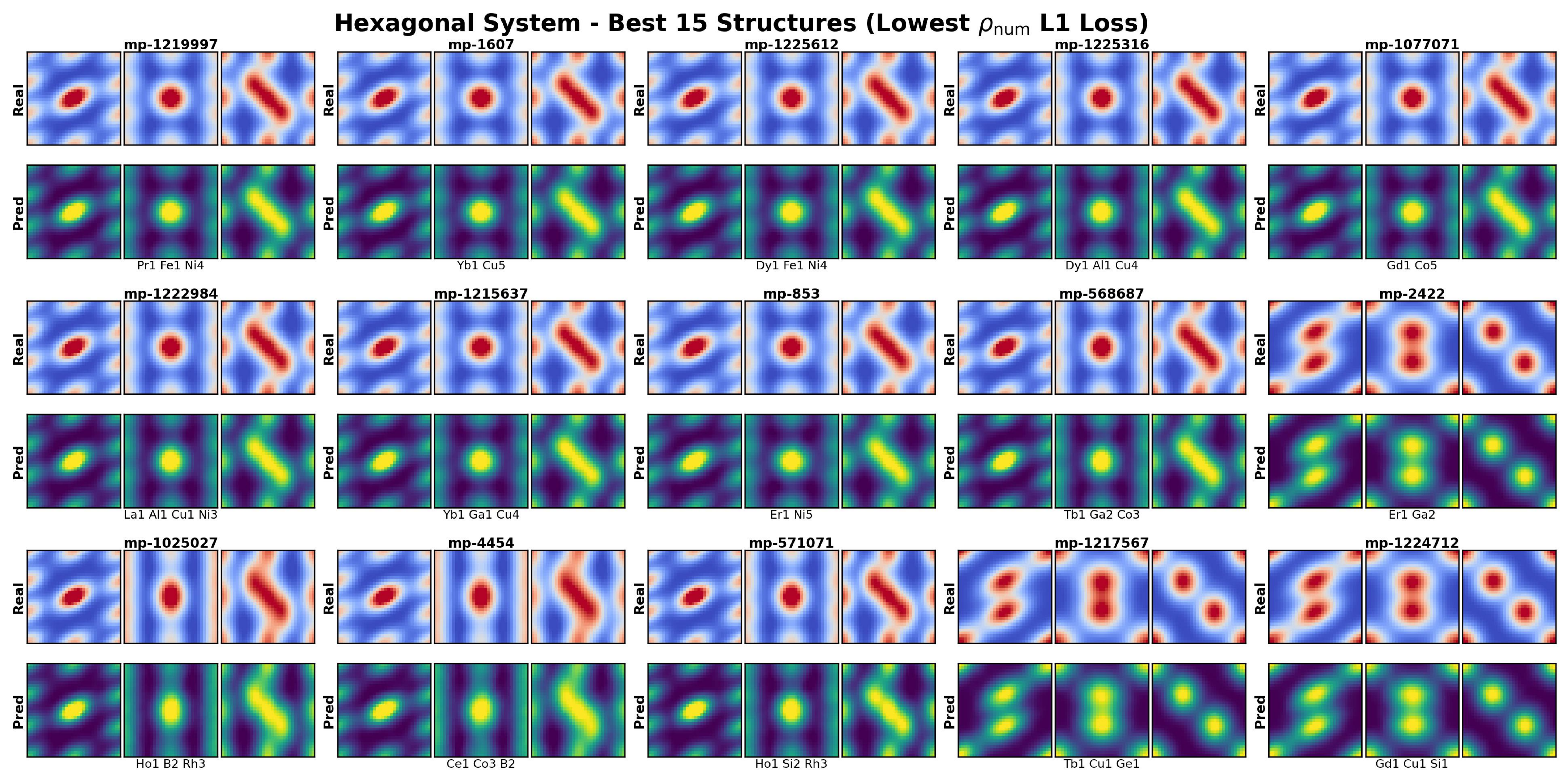}
\caption{Additional density field comparison for 15 representative hexagonal structures.
Top: ground truth; bottom: CrystalFNO Solver predictions.}
\label{fig:app_hexagonal}
\end{figure*}

\begin{figure*}[htbp]
\centering
\includegraphics[width=0.9\textwidth]{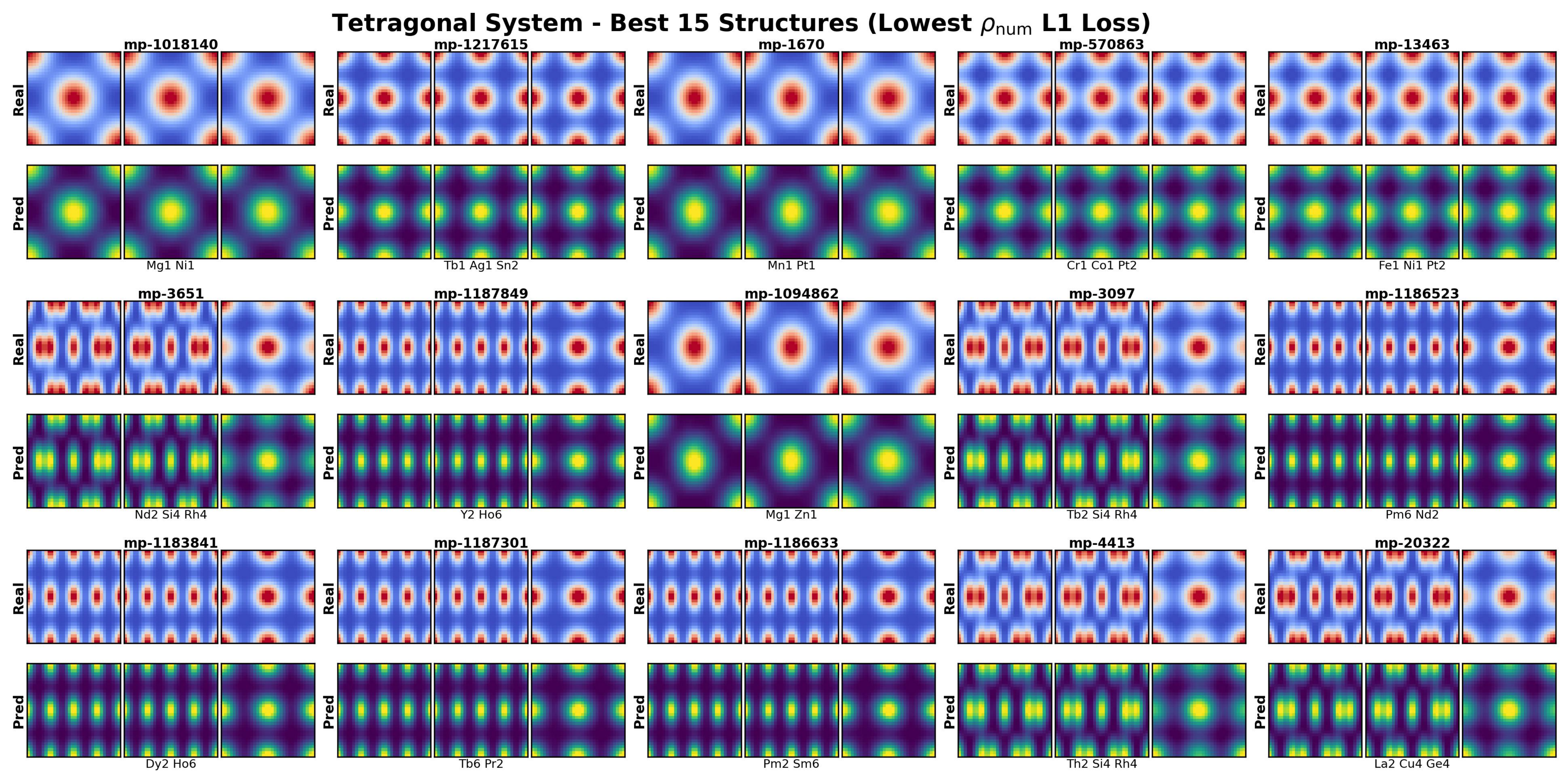}
\caption{Additional density field comparison for 15 representative tetragonal structures.
Top: ground truth; bottom: CrystalFNO Solver predictions.}
\label{fig:app_tetragonal}
\end{figure*}

\begin{figure*}[htbp]
\centering
\includegraphics[width=0.9\textwidth]{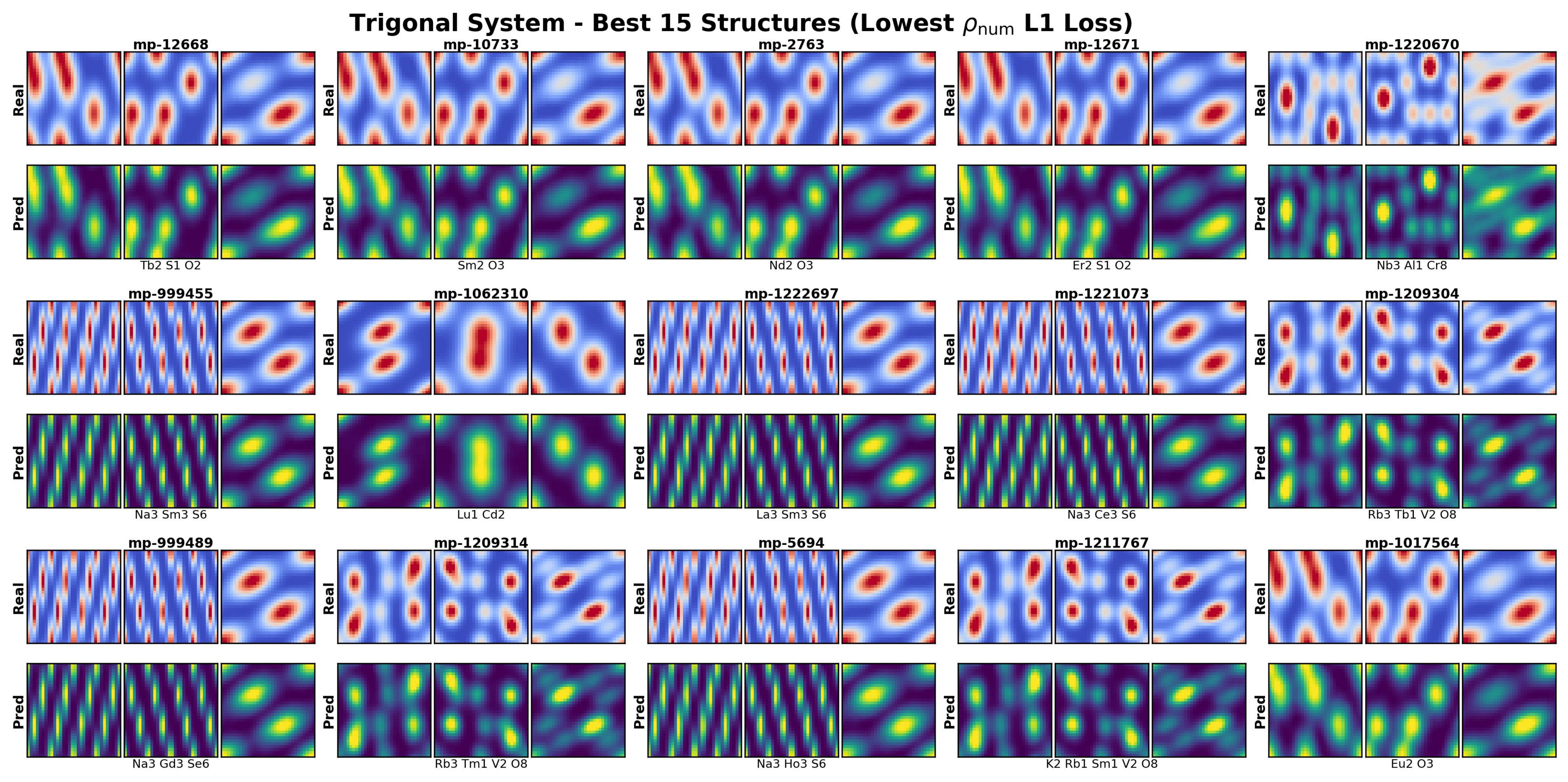}
\caption{Additional density field comparison for 15 representative trigonal structures.
Top: ground truth; bottom: CrystalFNO Solver predictions.}
\label{fig:app_trigonal}
\end{figure*}

\begin{figure*}[htbp]
\centering
\includegraphics[width=0.9\textwidth]{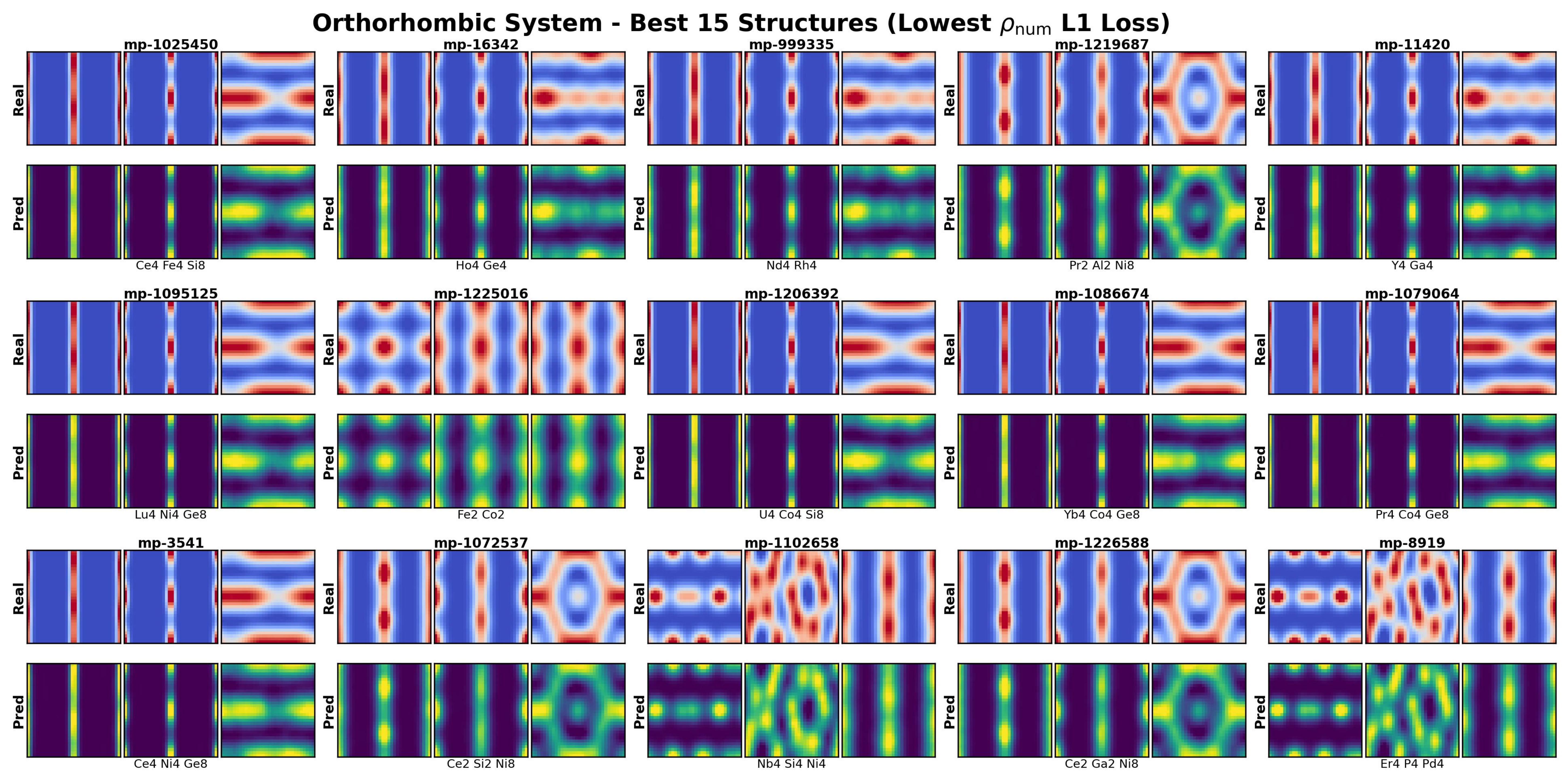}
\caption{Additional density field comparison for 15 representative orthorhombic
structures. Top: ground truth; bottom: CrystalFNO Solver predictions.}
\label{fig:app_orthorhombic}
\end{figure*}

\begin{figure*}[htbp]
\centering
\includegraphics[width=0.9\textwidth]{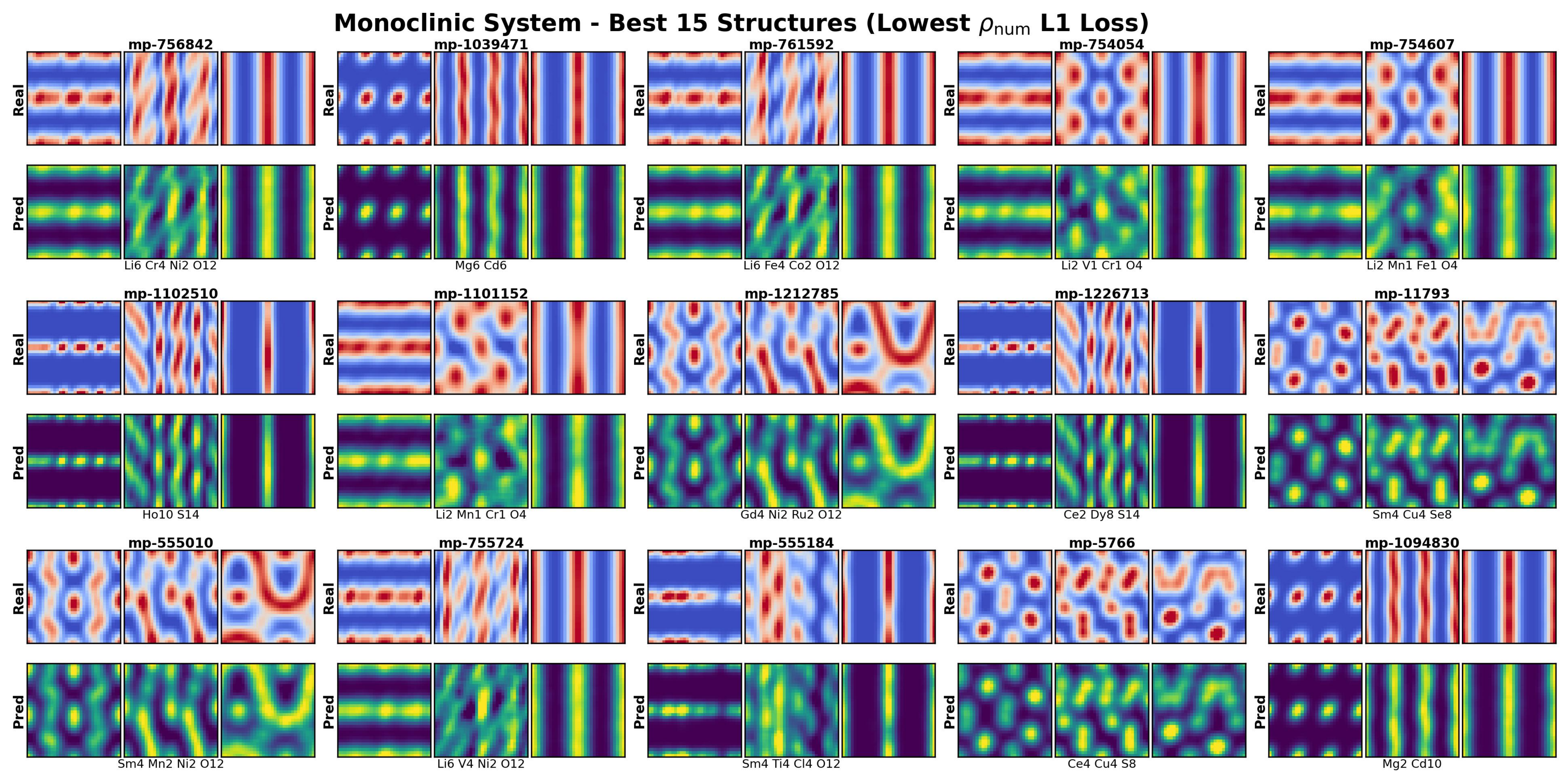}
\caption{Additional density field comparison for 15 representative monoclinic structures.
Top: ground truth; bottom: CrystalFNO Solver predictions.}
\label{fig:app_monoclinic}
\end{figure*}

\begin{figure*}[htbp]
\centering
\includegraphics[width=0.9\textwidth]{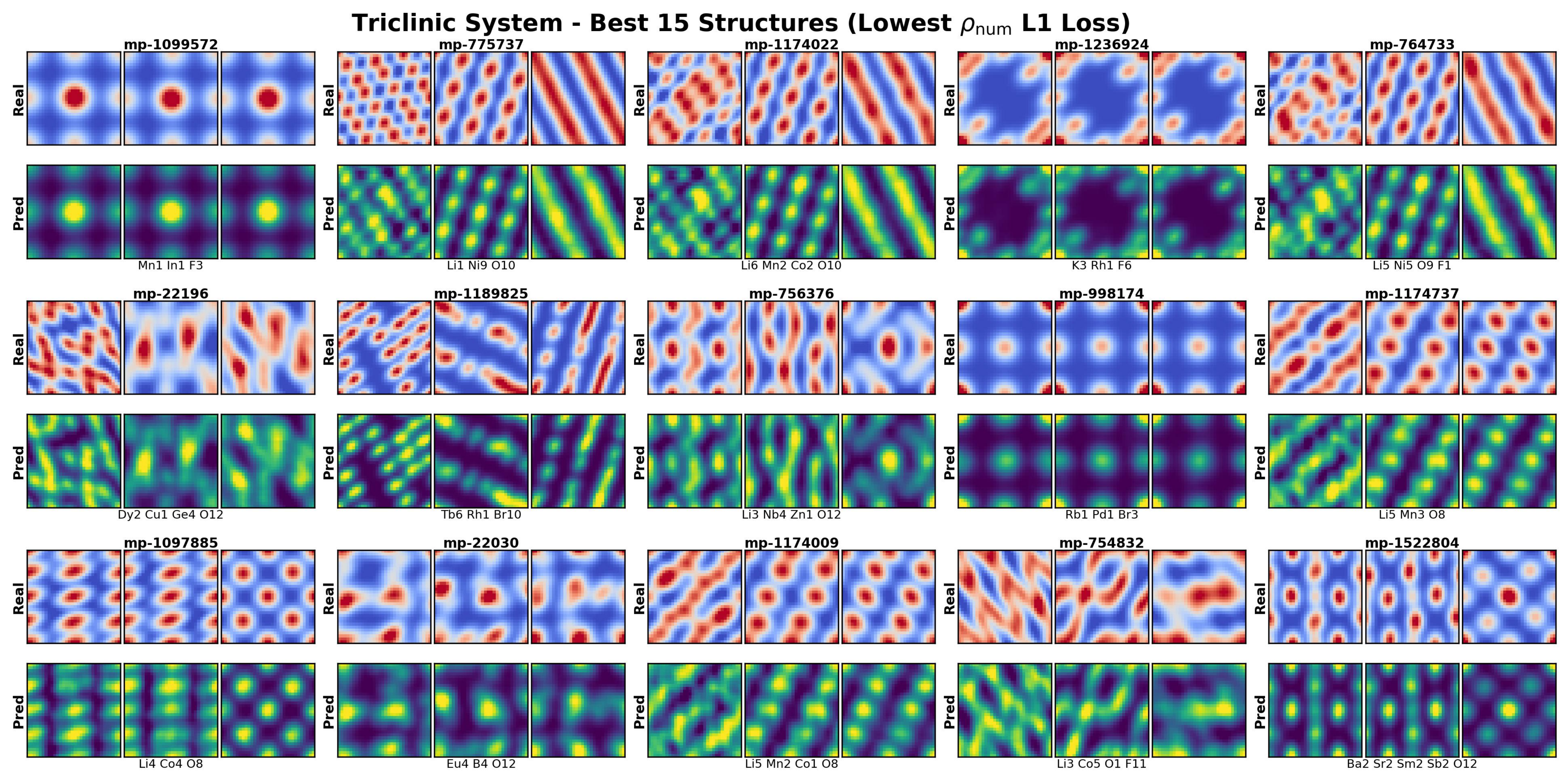}
\caption{Additional density field comparison for 15 representative triclinic structures.
Top: ground truth; bottom: CrystalFNO Solver predictions.}
\label{fig:app_triclinic}
\end{figure*}

We provide additional qualitative results for the CrystalFNO Solver density field prediction on the test set. For each of the seven crystal systems, we select 15 representative structures with low prediction errors and visualize the comparison between the ground-truth and predicted density fields. Figs.~\ref{fig:app_cubic}--\ref{fig:app_triclinic} present these results, with crystal systems ordered from cubic to triclinic. In each figure, the top row shows the ground-truth density fields and the bottom row shows the CrystalFNO Solver predictions, with three orthogonal projections (X, Y, Z) displayed for each structure.

\end{document}